\documentclass[12pt, leqno]{article}
\usepackage{oxford3}
\usepackage{kokoszka}
\usepackage{amssymb}
\usepackage{amsfonts}
\usepackage{epsfig}
\usepackage{amsmath}
\usepackage{enumitem}
\usepackage{color}

\DeclareMathOperator{\diag}{diag}
\DeclareMathOperator{\sd}{sd}

\renewcommand{\baselinestretch}{1.05}
\begin{document}

\title{Principal component analysis  of periodically correlated functional
time series}
\author{Lukasz Kidzi{\'n}ski \\
{\small Stanford University}
\and Piotr Kokoszka\thanks{Supported by NSF grant DMS 1462067.} \\
{\small Colorado State University} \and Neda Mohammadi Jouzdani \\
{\small Colorado State University }}
\date{}
\maketitle

\begin{abstract}
Within the framework of functional data analysis,
we develop principal component analysis for periodically correlated
time series of functions. We define the components of the above analysis
including periodic, operator--valued filters,  score processes and
the inversion formulas. We show that these objects are defined
via convergent series under a simple condition requiring
summability of the Hilbert--Schmidt norms of the filter coefficients,
and that they poses optimality properties.
We explain  how the Hilbert space theory reduces to an approximate
finite--dimensional setting which is implemented in a custom build
\verb|R| package. A data example and a simulation study
show that the new methodology is superior to existing tools if the
functional time series exhibit periodic characteristics.

\end{abstract}


\section{Introduction} \label{s:i} Periodicity is one of the most
important characteristics of time series, with early work going back
to the very origins of the field, e.g.\ \citetext{walker:1914} and
\citetext{fisher:1929}. The class of periodically correlated time
series is particularly suitable to quantify periodic behavior
reflected not only in the mean structure but also in correlations.
Consequently, periodically correlated (PC) time series have been used
in many modeling and prediction applications and various aspects of
their theory have been studied.  The book of
\citetext{hurd:miame:2007} gives an account of the subject.  It is
impossible to list even a fraction of relevant references, but to
indicate the many flavors of work done in this field we cite
\citetext{hurd:1989}, \citetext{lund:1995},
\citetext{anderson:meerschaert:1997}, \citetext{javorskyj:2012} and
\citetext{ghanbarzadeh:amingfari:2016}.

The last decade has seen  increased interest in time series of
curves, often referred to as functional time series (FTS's).
Examples of FTS's include annual temperature or smoothed precipitation
curves, e.g. \citetext{gromenko:kokoszka:reimherr:2016}, daily
pollution level curves, e.g.
\citetext{aue:dubartNorinho:hormann:2015}, various daily curves
derived from high frequency asset price data, e.g.
\citetext{horvath:kokoszka:rice:2014}, yield curves, e.g.
\citetext{hays:shen:huang:2012}, daily vehicle traffic curves, e.g.
\citetext{klepsch:kluppelberg:wei:2017}. Other examples are given in
the books of \citetext{HKbook} and \citetext{kokoszka:reimherr:2017}.
The theory and methodology of FTS's forms a subfield of Functional
Data  Analysis (FDA). A key tool of FDA is dimension reduction via
functional principal component analysis (FPCA), e.g. Chapter 3
of \citetext{HKbook}. FPCA has been developed for random samples
of functions, i.e for iid functional data. Recently,
\citetext{hormann:kidzinski:hallin:2015} extended the theory
of \citetext{brillinger:1975}, Chapter 9, from the setting
linear vector-valued time series to functional weakly dependent
time series.  Building on earlier advances of
 Panaretos and Tavakoli
(\citeyear{panaretos:tavakoli:2013AS},
\citeyear{panaretos:tavakoli:2013SPA}), they developed
spectral domain PCA which leads to a better representation
of stationary FTS's than the ususal (static) PCA. Suitable
details and definitions are given in Section~\ref{s:n}.
The objective of this paper is to develop PCA for periodically correlated
FTS's. We establish the requisite theoretical framework and show
that for FTS's with periodic characteristics the new approach  is superior
to the methodology of \citetext{hormann:kidzinski:hallin:2015}.
We emphasize that the latter methodology was developed for stationary
FTS's and so is a priori not well suited for periodic functional data.
Tests for periodicity in FTS's have recently been developed by
\citetext{hormann:kokoszka:nisol:2016} and
\citetext{zamani:2016}. \citetext{zhang:2016} uses spectral
methods to develop goodness of fit tests for FTS's.

Section~\ref{s:n} introduces the requisite background and notation. The
theory for the principal component analysis of periodically correlated
FTS's is presented in Section~\ref{s:DFPC}, with proofs postponed to
Section~\ref{s:p}. Section~\ref{s:ni} shows how the methodology
developed in the infinite dimensional framework of function spaces
is translated to an implementable setting of finite dimensional objects.
Its usefulness is illustrated in Section~\ref{s:a} by an application
to a particulate pollution data set and a simulation study.

\section{Notation and Preliminaries} \label{s:n}
This section introduces notation and background
used throughout the paper.  A generic
separable Hilbert space is denoted by $\mathbb{H}$, its inner product
and norm, respectively, by $<.,.>_{\mathbb{H}}$ and $\Vert .\Vert
_{\mathbb{H}}$.  The subscript  $\mathbb{H}$ is sometimes suppressed
when there is no ambiguity.

The
Hilbert space $\mathcal{H}=L^{2}\left( \left[ 0,1\right] \right) $ and its $%
T $-fold Cartesian product $\mathcal{H}^{T}$ are extensively used throughout
the paper. They are equipped with inner products
\begin{equation*}
\left\langle f,g\right\rangle _{\mathcal{H}}=\int_{0}^{1}f\left( s\right)
\overline{g}\left( s\right) ds,\text{ \ \ \ }f,g\in \mathcal{H}
\end{equation*}
and
\begin{equation*}
\left\langle \left(
\begin{array}{ccc}
f_{1} & \cdots & f_{T}
\end{array}
\right) ^{^{\prime }},\left(
\begin{array}{ccc}
g_{1} & \cdots & g_{T}
\end{array}
\right) ^{^{\prime }}\right\rangle _{\mathcal{H}^{T}}=\sum_{j=1}^{T}\left
\langle f_{j},g_{j}\right\rangle _{\mathcal{H}},\text{\ \ \ \ }
f_{j},g_{j}\in \mathcal{H},
\end{equation*}
respectively.  An operator $\Psi $ from a Hilbert space $\mathbb{H}$
to $\mathbb{C}^{p}$ is a bounded linear operator if and only if there
exist (unique) elements $%
\Psi _{1},\ldots ,\Psi _{p}$ in $\mathbb{H}$ such that
\begin{equation}  \label{Representers}
\Psi \left( h\right)
=\left( \left\langle h,\Psi _{1}\right\rangle _{\mathbb{H}},\ldots ,
\left\langle h,\Psi _{p}\right\rangle _{\mathbb{H}}\right)
^{^{\prime }},\text{ \ \ \ }\forall\ h\in \mathbb{H}.
\end{equation}
An operator $\Upsilon $ from $\mathbb{C}^{p}$\ to $\mathbb{H}$ is
linear and bounded if and only if there exist elements $\Upsilon
_{1},\ldots ,\Upsilon _{p}$ in $\mathbb{H}$ such that
\begin{equation*}
\Upsilon \left( y\right) =\Upsilon \left( \left( y_{1},\ldots ,y_{p}\right)
^{^{\prime }}\right) =\sum_{m=1}^{p}y_{m}\Upsilon _{m}, \text{ \ \ \ }
\forall \ y\in \mathbb{C}^{p}.
\end{equation*}
For any two elements $f$\ and $g$\ in $\mathbb{H}$, $f\otimes g$\ is a
bounded linear operator defined by
\begin{equation*}
f\otimes g : \mathbb{H\longrightarrow H}, \ \ \ f\otimes g : h \longmapsto
\left\langle h,g\right\rangle _{\mathbb{H}}f.
\end{equation*}
We use $\left\Vert .\right\Vert_{\mathcal{L}}$ to denote the operator norm,
and $\left\Vert .\right\Vert _{\mathcal{N}}$ and $\left\Vert .\right\Vert _{
\mathcal{S}}$ to denote, respectively,
 the nuclear and Hilbert-Schmidt norms, e.g. \citetext{HKbook},
 Section 13.5.

In the following $L^{2}\left( \mathbb{H},\left( -\pi ,\pi \right] \right) $
denotes the space of square integrable $\mathbb{H}$-valued functions on $%
\left( -\pi ,\pi \right] $. Similarly, for a probability  space $
\left( \Omega ,\mathcal{A},\mathbb{P}\right) $\ in place of $\left( -\pi
,\pi \right] $, we use the notation $L^{2}\left( \mathbb{H},\Omega
\right) $.  For two random elements $X,Y\in L^{2}\left(
\mathbb{H},\Omega \right) $, the covariance operator,
$\mathrm{Cov}\left( X,Y\right) $, is defined as
\begin{eqnarray*}
\mathrm{Cov}\left( X,Y\right) &=&E\left[ \left( X-EX\right) \otimes \left(
Y-EY\right) \right] :\mathbb{H\longrightarrow H}, \\
\mathrm{Cov}\left( X,Y\right) &:&h\longmapsto E\left[ \left\langle h,\left(
Y-EY\right) \right\rangle _{\mathbb{H}}\left( X-EX\right) \right] .
\end{eqnarray*}

\begin{definition}
Let $X=\left\{ X_{t},t\in \mathbb{Z}\right\} $ be an $\mathbb{H}$-valued
time series
with finite second moment, $E\| X_t \|^2 < \infty$. Then,  $X$  is said to be
 periodically correlated if there exists a positive integer $T$
such that
\begin{eqnarray*}
EX_{t} &=&EX_{t+T},\text{ \ \ }\forall \ t\in \mathbb{Z},  \\
\mathrm{Cov}\left( X_{t},X_{s}\right) &=&\mathrm{Cov}\left(
X_{t+T},X_{s+T}\right) ,\text{\ \ \ }\forall \ t,s\in \mathbb{Z}.
\end{eqnarray*}
The smallest such $T$ will be called the period of the process, and
$X$\ is then said to be $T$-periodically correlated, or $T$-PC,  for
short. When $T=1$, the process is (weakly) stationary.
\end{definition}

For a $T$-PC\ process $X$, covariance operators at lag $h$ are defined as
\begin{equation*}
C_{h,\left( j,j^\prime \right) }^{X}=\mathrm{Cov}\left(
X_{Th+j},X_{j^\prime}\right) ,\text{ \ \ \ }h\in \mathbb{Z}\text{ and }%
j,j^\prime=0,1,\ldots ,T-1.
\end{equation*}%
It is easy to verify that the condition
\begin{equation}
\sum_{h\in \mathbb{Z}}\left\Vert C_{h,\left( j,j^\prime\right)
}^{X}\right\Vert _{\mathcal{S}}<\infty ,\text{ \ \ \ }j,j^{^{\prime
}}=0,1,\ldots ,T-1,  \label{HS norm summable}
\end{equation}%
implies that for each $\theta $ the series $\left\{ \frac{1}{2\pi }\sum_{h=%
\mathbb{-}n}^{n}C_{h,\left( j,j^\prime\right) }^{X}e^{-ih\theta }:n\in
\mathbb{Z}_{+}\right\} $\ is a Cauchy\ sequence in the Hilbert space of
Hilbert-Schmidt operators on $\mathbb{H}$. Then, spectral density operators
are well defined by
\begin{equation}
\mathcal{F}_{\theta ,\left( j,j^\prime\right) }^{X}=\frac{1}{2\pi }%
\sum_{h\in \mathbb{Z}}C_{h,\left( j,j^\prime\right) }^{X}e^{-ih\theta
},\text{ \ \ \ }j,j^\prime=0,\ldots ,T-1.  \label{SDO components}
\end{equation}

\begin{remark}
\label{finiteness of E||X_t||^2}
The periodic behavior of covariance operator of $X$
implies that the set $\left\{ E\left\Vert X_{t}\right\Vert
^{2},t\in \mathbb{Z}\right\} $ is finite.
It consists of at most $T$ elements because
\begin{equation*}
E\left\Vert X_{t}\right\Vert ^{2}={\rm Tr}\left( \mathrm{Cov}\left(
X_{t},X_{t}\right) \right) ={\rm Tr}\left( \mathrm{Cov}\left(
X_{t+T},X_{t+T}\right) \right).
\end{equation*}
\end{remark}

\begin{definition}
A sequence $\left\{ \Psi _{l},l\in \mathbb{Z}\right\} $ of operators from a
Hilbert space $\mathbb{H}_{1}$\ to a Hilbert space $\mathbb{H}_{2}$
\ satisfying
\begin{equation}
\sum_{l\in \mathbb{Z}}\left\Vert \Psi _{l}\right\Vert _{\mathcal L}<\infty ,
\label{filter coef}
\end{equation}
is called a filter. A $T$-periodic filter $\left\{ \left\{ \Psi _{l}^{t},l\in
\mathbb{Z}\right\} ,t\in \mathbb{Z}\right\} $\ is a sequence of filters
which is $T$-periodic with respect to $t$ i.e.
$\Psi _{l}^{t}=\Psi _{l}^{t+T}$, for each $t$ and $l$. Consequently,
\begin{equation} \label{PC filter coef}
\sum_{t=0}^{T-1}\sum_{l\in \mathbb{Z}}\left\Vert \Psi _{l}^{t}\right\Vert
_{\mathcal L}<\infty .
\end{equation}
\end{definition}
Related to the filter $\left\{ \Psi _{l},l\in \mathbb{Z}\right\} $, $\Psi
\left( B\right) $\ is an operator from $\left( \mathbb{H}_{1}\right) ^{%
\mathbb{Z}}$\ to $\left( \mathbb{H}_{2}\right) ^{\mathbb{Z}}$\ of the
following form
\begin{equation*}
\Psi \left( B\right) =\sum_{l\in \mathbb{Z}}\Psi _{l}B^{l},
\end{equation*}
where $B$\ is the backward shift operator. In the other words, if $\left\{
X_{t},t\in \mathbb{Z}\right\} $ is a time series with values in $\mathbb{H}%
_{1}$, then $\Psi \left( B\right) $ transforms it to an $\mathbb{H}_{2}$
-valued time series defined by
\begin{equation*}
\left( \Psi \left( B\right) \left( X\right) \right) _{t}
=\sum_{l\in \mathbb{Z}}\Psi _{l}\left( X_{t-l}\right).
\end{equation*}

For a $p\times p$\ matrix $\mathbf{A}$, $a_{q,r}$ denotes the entry in
the $ q$-th row and $r$-th column. To indicate that $t= kT+ d$ for
some integer $k$, we write $t\overset{T}{ \equiv }d$.

\section{Principal component analysis of periodically correlated
functional time series}
\label{s:DFPC} Before proceeding with the definitions and statements of
properties of the principal component analysis
for PC-FTS's, we provide a brief introduction, focusing on the ideas and
omitting  mathematical assumptions. Suppose $\lbr X_{t}\rbr$ is a weakly
dependent, stationary mean zero time series of functions in $\mathcal{H}$.
It admits the Karhunen--Lo{\'{e}}ve expansion
\begin{equation}
X_{t}(u)=\sum_{m=1}^{\infty }\xi _{tm}v_{m}(u),\ \ \
E\xi _{tm}^{2}=\la_{m},
\label{e:KL}
\end{equation}
where the $v_m$ are the functional principal components (called \emph{
static} FPC's in \citetext{hormann:kidzinski:hallin:2015}). The
orthonormal functions $v_{m}$ are uniquely defined up to a sign, and
the random variables $\xi _{tm}$ are called their scores. Even for
stationary (rather than periodically correlated) functional time
series, the \emph{dynamic} FPC's are not defined as one function for
every ``frequency'' level $m$. The analog of \refeq{KL} is
\begin{equation}
X_{t}(u)=\sum_{m=1}^{\infty }\sum_{l\in {\mathbb{Z}}}Y_{m,t+l}\phi _{ml}(u).
\label{e:KLD}
\end{equation}
A single function $v_{m}$ is thus replaced by an infinite sequence of
functions $\lbr\phi _{ml},l\in {\mathbb{Z}}\rbr$. However, one can still
define the scores as single numbers for every frequency level $m$, using the
formula $Y_{mt}=\sum_{l\in {\mathbb{Z}}}\lip X_{t-l},\phi _{ml}\rip$. The
analog of $\la_{m}$ is
\begin{equation*}
\nu _{m}:=E\lnorm\sum_{l\in {\mathbb{Z}}}Y_{m,t+l}\phi _{ml}\rnorm^{2},
\end{equation*}
and we have the decomposition of variance $E\lnorm X_{t}\rnorm%
^{2}=\sum_{m=1}^{\infty }\nu _{m}$. In this section, we will see how
these results extend to the setting of periodically correlated
functional time series, which is necessarily more complex as it
involves periodic sequences of functions. The scores, and
reconstructions obtained from them, will have certain periodic
properties. \emph{All results stated in this section are proven in
  Section~\ref{s:p}. }

In order to define the dynamic functional principal components in our
setting, we first establish conditions for the existence of a filtered
(output) process of a $T $-PC functional times series. The periodic
structure of the covariance operators of the $T$-PC input process $X=\left\{
X_{t},t\in \mathbb{Z} \right\} $ suggests applying a $T$-periodic functional
filter $\left\{ \left\{ \Psi _{l}^{t},l\in \mathbb{Z}\right\} ,t\in \mathbb{Z
}\right\} $\ to obtain a filtered process $\mathbf{Y}=\left\{ \mathbf{Y}
_{t},t\in \mathbb{Z}\right\}$ with values in $\mathbb{C}^{p}$.

\begin{theorem} \label{T filter process} Let $X=\left\{ X_{t},t\in
    \mathbb{Z}\right\} $\ be an $\mathcal{H}$-valued $T$-PC process
  and $\left\{ \left\{ \Psi _{l}^{t},l\in \mathbb{Z}\right\} ,t\in
    \mathbb{Z}\right\} $ a $T$-periodic filter from $ \mathcal{H}$\ to
  $\mathbb{C}^{p}$\ with the  elements $\Psi _{l,m}^{t},m=1, \ldots ,p$
  in $\mathcal{H}$, as described in \eqref{Representers}.  In
  particular, we assume that \eqref{PC filter coef} holds.  Then, for
  each $t$, $\sum_{l\in \mathbb{Z}}\Psi _{l}^{t}\left( X_{\left(
        t-l\right) }\right) $ converges in mean square to a limit
  $\mathbf{Y}_{t}$.

If,  in addition,
\begin{equation}
\sum_{t=0}^{T-1}\sum_{l\in \mathbb{Z}}\left\Vert \Psi _{l}^{t}\right\Vert _{
\mathcal{S}}<\infty ,  \label{S summable}
\end{equation}
then $\mathbf{Y}=\left\{ \mathbf{Y}_{t},t\in \mathbb{Z}\right\} $ is a
$T$-PC process with the following $p\times p$  spectral density
matrices $\mathcal{F}_{\theta,\left( d,f\right) }^{\mathbf{Y}}$ for
$d,f=0, \ldots ,T-1, $
\begin{align*}
&\mathcal{F}_{\theta,\left( d,f\right) }^{\mathbf{Y}}\\
&\ \  =\left[\left\langle \left(
\begin{array}{ccc}
\mathcal{F}_{\theta,\left( 0,0\right) }^{X} & \cdots  & \mathcal{F}_{\theta
,\left( 0,T-1\right) }^{X} \\
\vdots  & \ddots  & \vdots  \\
\mathcal{F}_{\theta,\left( T-1,0\right) }^{X} & \cdots  & \mathcal{F}
_{\theta,\left( T-1,T-1\right) }^{X}
\end{array}
\right) \left(
\begin{array}{c}
\Psi _{\theta,d,r}^{d} \\
\vdots  \\
\Psi _{\theta,d-T+1,r}^{d}
\end{array}
\right),\left(
\begin{array}{c}
\Psi _{\theta,f,q}^{f} \\
\vdots  \\
\Psi _{\theta,f-T+1,q}^{f}
\end{array}
\right) \right\rangle _{\mathcal{H}^{T}}\right]_{q,r=1,\ldots ,p}
\end{align*}
where $\Psi _{\theta,d,q}^{d}=\sum_{l\in \mathbb{Z}}\Psi _{Tl+d,q}^{d}e^{il\theta
},\ldots,\Psi _{\theta,d-T+1,q}^{d}=\sum_{l\in \mathbb{Z}}\Psi
_{Tl+d-T+1,q}^{d}e^{il\theta }, f,d=0,\ldots,T-1$.
\end{theorem}

To illustrate the spectral density structure of the output process, we
consider $T=2$, in which case,
\begin{equation*}
\mathcal{F}_{\theta ,\left( 0,0\right) }^{\mathbf{Y}}=\left[ \left\langle
\left (
\begin{array}{cc}
\mathcal{F}_{\theta ,\left( 0,0\right) }^{X} & \mathcal{F}_{\theta ,\left(
0,1\right) }^{X} \\
\mathcal{F}_{\theta ,\left( 1,0\right) }^{X} & \mathcal{F}_{\theta ,\left(
1,1\right) }^{X}
\end{array}
\right) \left(
\begin{array}{c}
\Psi _{\theta ,0,r}^{0} \\
\Psi _{\theta ,-1,r}^{0}
\end{array}
\right) ,\left(
\begin{array}{c}
\Psi _{\theta ,0,q}^{0} \\
\Psi _{\theta ,-1,q}^{0}
\end{array}
\right) \right\rangle _{\mathcal{H}^{2}}\right] _{q,r=1,\ldots ,p},
\end{equation*}
\begin{equation*}
\mathcal{F}_{\theta ,\left( 1,0\right) }^{\mathbf{Y}}=\left[ \left\langle
\left(
\begin{array}{cc}
\mathcal{F}_{\theta ,\left( 0,0\right) }^{X} & \mathcal{F}_{\theta ,\left(
0,1\right) }^{X} \\
\mathcal{F}_{\theta ,\left( 1,0\right) }^{X} & \mathcal{F}_{\theta ,\left(
1,1\right) }^{X}
\end{array}
\right) \left(
\begin{array}{c}
\Psi _{\theta ,1,r}^{1} \\
\Psi _{\theta ,0,r}^{1}
\end{array}
\right) ,\left(
\begin{array}{c}
\Psi _{\theta ,0,q}^{0} \\
\Psi _{\theta ,-1,q}^{0}
\end{array}
\right) \right\rangle _{\mathcal{H}^{2}}\right]_{q,r=1,\ldots ,p},
\end{equation*}
\begin{equation*}
\mathcal{F}_{\theta ,\left( 0,1\right) }^{\mathbf{Y}}=\left[ \left\langle
\left(
\begin{array}{cc}
\mathcal{F}_{\theta ,\left( 0,0\right) }^{X} & \mathcal{F}_{\theta ,\left(
0,1\right) }^{X} \\
\mathcal{F}_{\theta ,\left( 1,0\right) }^{X} & \mathcal{F}_{\theta ,\left(
1,1\right) }^{X}
\end{array}
\right) \left(
\begin{array}{c}
\Psi _{\theta ,0,r}^{0} \\
\Psi _{\theta ,-1,r}^{0}
\end{array}
\right) ,\left(
\begin{array}{c}
\Psi _{\theta ,1,q}^{1} \\
\Psi _{\theta ,0,q}^{1}
\end{array}
\right) \right\rangle _{\mathcal{H}^{2}}\right]_{q,r=1,\ldots ,p},
\end{equation*}
\begin{equation*}
\mathcal{F}_{\theta ,\left( 1,1\right) }^{\mathbf{Y}}
=\left[ \left\langle \left(
\begin{array}{cc}
\mathcal{F}_{\theta ,\left( 0,0\right) }^{X} & \mathcal{F}_{\theta ,\left(
0,1\right) }^{X} \\
\mathcal{F}_{\theta ,\left( 1,0\right) }^{X} & \mathcal{F}_{\theta ,\left(
1,1\right) }^{X}
\end{array}
\right) \left(
\begin{array}{c}
\Psi _{\theta ,1,r}^{1} \\
\Psi _{\theta ,0,r}^{1}
\end{array}
\right) ,\left(
\begin{array}{c}
\Psi _{\theta ,1,q}^{1} \\
\Psi _{\theta ,0,q}^{1}
\end{array}
\right) \right\rangle _{\mathcal{H}^{2}}\right]_{q,r=1,\ldots ,p},
\end{equation*}
where $\Psi _{\theta ,0,q}^{0}:=\sum_{l\in \mathbb{Z}}\Psi
_{2l,q}^{0}e^{il\theta }$, $\Psi _{\theta ,-1,q}^{0}:=\sum_{l\in \mathbb{Z}
}\Psi _{2l-1,q}^{0}e^{il\theta }$, $\Psi _{\theta ,0,q}^{1}:=\sum_{l\in
\mathbb{Z}}\Psi _{2l,q}^{1}e^{il\theta }$, and $\Psi _{\theta
,1,q}^{1}:=\sum_{l\in \mathbb{Z}}\Psi _{2l+1,q}^{1}e^{il\theta }$.

We emphasize that \eqref{PC filter coef} is a sufficient condition for
the mean-square convergence of the series defining the filtered
process $\mathbf{Y}$, and \eqref{S summable} guarantees the existence
of spectral density operator of the filtered
process. \citetext{hormann:kidzinski:hallin:2015}, page 327, discuss
this issue in the case of stationary input and output
processes. \emph{In the remainder of the paper, we assume \eqref{S
summable} for each periodic functional filter.}

The operator matrix $\left( \mathcal{F}_{\theta ,\left( d, f \right)
}^{X}\right) _{0\leq d, f\leq T-1}$ in Theorem \ref{T filter process} is a
non-negative, self-adjoint compact operator from $\mathcal{H}^{T}$\ to $
\mathcal{H}^{T}$, and so it admits the following spectral decomposition
\begin{equation}  \label{eigenvf decomposition}
\left( \mathcal{F}_{\theta ,\left( d, f \right) }^{X}\right) _{0\leq d, f\leq
T-1} =\sum_{m\geq 1}\lambda _{\theta ,m}\varphi _{\theta ,m}\otimes \varphi
_{\theta ,m},
\end{equation}
where $\lambda _{\theta ,1}\geq \lambda _{\theta ,2}\geq \cdots \geq 0$, and
$\left\{ \varphi _{\theta ,m}\right\} _{m\geq 1}$ forms a complete
orthonormal basis for $\mathcal{H}^{T}$. By choosing $\left(
\begin{array}{ccc}
\Psi _{\theta ,d,q}^{d} & \cdots & \Psi _{\theta ,d-T+1,q}^{d}
\end{array}
\right) ^\prime$ as the eigenfunction $\varphi _{\theta ,dp+q}$, the
spectral density matrices of the filtered process $\mathbf{Y}=\left\{
\mathbf{Y}_{t},t\in \mathbb{Z}\right\} $\ turn to diagonal matrices and an
optimality property will be obtained. We are now ready to define the DFPC
filter and scores of the periodically correlated process $X$.

\begin{definition}
Let $X=\left\{ X_{t},t\in \mathbb{Z}\right\} $ be an $\mathcal{H}$
-valued mean zero $T$-PC\ random process satisfying condition
\eqref{HS norm summable}
and $\left\{ \Phi _{l,m}^{d},d=0,\ldots,T-1,m=1,\ldots p,l\in \mathbb{Z}\right\} $ be
elements of $\mathcal{H}$ defined by
\begin{equation}
\dfrac{1}{2\pi}\int_{-\pi }^{\pi }\varphi _{\theta ,dp+m}e^{-il\theta }d\theta =\left(
\begin{array}{c}
\Phi _{lT+d,m}^{d} \\
\vdots\\
\Phi _{lT+d-T+1,m}^{d}
\end{array}
\right)
 ,\text{\ \ \ \ }m=1,\ldots ,p,\text{\ }d=0,\ldots,T-1  \label{T DF filter}
\end{equation}
for each $l$ in $\mathbb{Z}$,  or equivalently by
\begin{equation}
\varphi _{\theta ,dp+m}=\left(
\begin{array}{c}
\Phi _{\theta ,d,m}^{d} \\
\vdots\\
\Phi _{\theta ,d-T+1,m}^{d}
\end{array}
\right)
,\text{\ \ \ \ }m=1,\ldots ,p,\text{\ }d=0,\ldots,T-1,  \label{T DFPC}
\end{equation}
for each $\theta $\ in $\left( -\pi ,\pi \right] $. Then,
\begin{equation*}
\left\{ \Phi _{l,m}^{d},l\in \mathbb{Z}\right\} ,\text{\ \ \ \ }d=0,\ldots,T-1,
\end{equation*}
is said to be the $(d,m)$-th dynamic functional principal component (DFPC)
filter of the process $X$. Furthermore,
\begin{eqnarray}\label{e:scores}
Y_{t,m}&=&\sum_{l\in \mathbb{Z}}\left\langle X_{\left( t-l\right) },\Phi
_{l,m}^{d}\right\rangle\\
&=&\sum_{l\in \mathbb{Z}}\left\langle X_{\left( t-lT-d\right) },\Phi
_{lT+d,m}^{d}\right\rangle+\sum_{l\in \mathbb{Z}}\left\langle X_{\left( t-lT-d+1\right) },\Phi
_{lT+d-1,m}^{d}\right\rangle\nn\\
&&+\cdots+\sum_{l\in \mathbb{Z}}\left\langle X_{\left( t-lT-d+T-1\right) },\Phi
_{lT+d-T+1,m}^{d}\right\rangle
, \text{\ \ \ \ } m=1, \ldots p,\text{\ } t\overset{T}{\equiv}d \nn
\end{eqnarray}
will be called the $(t,m)$-th DFPC score of $X$.
\end{definition}

For illustration, in case of $T=2$, we have for $m=1,\ldots ,p$,
\begin{equation}
\dfrac{1}{2\pi }\int_{-\pi }^{\pi }\varphi _{\theta ,m}e^{-il\theta }d\theta
=\left(
\begin{array}{c}
\Phi _{2l,m}^{0} \\
\Phi _{2l-1,m}^{0}
\end{array}
\right) \text{\ }\mathrm{and}\text{ }\dfrac{1}{2\pi }\int_{-\pi }^{\pi
}\varphi _{\theta ,p+m}e^{-il\theta }d\theta =\left(
\begin{array}{c}
\Phi _{2l+1,m}^{1} \\
\Phi _{2l,m}^{1}
\end{array}
\right) ,  \label{DF filter}
\end{equation}
for each $l$ in $\mathbb{Z}$, or equivalently,
\begin{equation*}
\varphi _{\theta ,m}=\left(
\begin{array}{c}
\Phi _{\theta ,0,m}^{0} \\
\Phi _{\theta ,-1,m}^{0}
\end{array}
\right) \text{\ }\mathrm{and}\text{ }\varphi _{\theta ,p+m}=\left(
\begin{array}{c}
\Phi _{\theta ,1,m}^{1} \\
\Phi _{\theta ,0,m}^{1}
\end{array}
\right) ,\ \ \ \theta \in \left( -\pi ,\pi \right] .
\end{equation*}
The filters $\left\{ \Phi _{l,m}^{d},l\in \mathbb{Z}\right\} $ are defined
for $d=0,1$.

The following proposition lists some useful properties of the
$p$-dimensional output process $\left\{ \mathbf{Y}_t = (Y_{t, 1},
\ldots, Y_{t,p} )^\prime, \ t \in \mathbb{Z}\right\}$ defined by
\eqref{e:scores}.

\begin{proposition}\label{p:properties}
Let $X=\left\{
X_{t},t\in \mathbb{Z}\right\} $ be an $\mathcal{H}$-valued mean zero
$T$-PC random process and assume that \eqref{HS norm summable}
holds. Then,
\begin{enumerate}[label=(\alph*)]
\item the eigenfunctions $\varphi _{m}\left( \theta \right) $\
are Hermitian i.e. $\varphi _{-\theta ,m}=\overline{\varphi} _{\theta
,m}$ and the DFPC scores $Y_{t,m}$\ are real-valued provided that $X$
is real-valued;
\item
 for each $(t,m)$, the series \eqref{e:scores}
 is mean-square convergent, has mean zero:
\begin{equation}
EY_{t,m}=0,  \label{EY}
\end{equation}
and satisfies for $t\overset{T}{\equiv }d$,
\begin{eqnarray}\label{T E||Y||^2 ;t=00}
E\left\Vert Y_{t,m}\right\Vert
^{2}
&=\sum\limits_{j_{1},j_{2}=0}^{T-1}\sum\limits_{k,l\in \mathbb{Z}
}\left\langle C_{k-l,(j_{1},j_{2})}^{X}\left( \Phi
_{kT+d-j_{2},m}^{d}\right),\Phi _{lT+d-j_{1},m}^{d}\right\rangle _{\mathcal{H}};
\end{eqnarray}
\item
 for any $t$ and $s$, the DFPC\ scores $Y_{t,m}$\ and
 $Y_{s,m^\prime}$ are uncorrelated if  $ s-t$ is not a multiple of $T$ or $m\neq m^\prime$. In other words
 $C_{h,\left(j_{1},j_{2}\right)}^{\mathbf{Y}}=0$ for $j_{1} \neq
 j_{2}$ and $ C_{h,(j,j)}^{\mathbf{Y}}$ are diagonal matrices for all
 $h$;
\item the long-run covariance matrix of the filtered process $\left\{
\mathbf{Y}_{t},t\in \mathbb{Z}\right\} $ satisfies the following limiting equality
\begin{equation*}
\lim_{n\rightarrow \infty }\frac{1}{n}\mathrm{Var}\left( \mathbf{Y}_{1}+\cdots
+\mathbf{Y}_{n}\right) =\dfrac{2\pi }{T}\sum_{d=0}^{T-1}\mathrm{diag}\left( \lambda _{0,dp+1}
,\ldots,\lambda _{0,dp+p} \right)  .
\end{equation*}
\end{enumerate}
\end{proposition}

For illustration, if $T=2$, then
\begin{eqnarray}
E\left\Vert Y_{t,m}\right\Vert ^{2} &=&\sum_{k\in \mathbb{Z}}\sum_{l\in
\mathbb{Z}}\left\langle C_{k-l,(0,0)}^{X}\left( \Phi _{2k,m}^{0}\right)
,\Phi _{2l,m}^{0}\right\rangle _{\mathcal{H}}  \label{E||Y||^2 ;t=00} \\
&&+\sum_{k\in \mathbb{Z}}\sum_{l\in \mathbb{Z}}\left\langle
C_{k-l,(0,1)}^{X}\left( \Phi _{2k-1,m}^{0}\right) ,\Phi
_{2l,m}^{0}\right\rangle _{\mathcal{H}}  \notag \\
&&+\sum_{k\in \mathbb{Z}}\sum_{l\in \mathbb{Z}}\left\langle
C_{k-l,(1,0)}^{X}\left( \Phi _{2k,m}^{0}\right) ,\Phi
_{2l-1,m}^{0}\right\rangle _{\mathcal{H}}  \notag \\
&&+\sum_{k\in \mathbb{Z}}\sum_{l\in \mathbb{Z}}\left\langle
C_{k-l,(1,1)}^{X}\left( \Phi _{2k-1,m}^{0}\right) ,\Phi
_{2l-1,m}^{0}\right\rangle _{\mathcal{H}},\text{ \ \ \ \ }t\overset{2}{
\equiv }0\nn
\end{eqnarray}
and
\begin{eqnarray}
E\left\Vert Y_{t,m}\right\Vert ^{2} &=&\sum_{k\in \mathbb{Z}}\sum_{l\in
\mathbb{Z}}\left\langle C_{k-l,(1,1)}^{X}\left( \Phi _{2k,m}^{1}\right)
,\Phi _{2l,m}^{1}\right\rangle _{\mathcal{H}}  \label{E||Y||^2 ;t=11} \\
&&+\sum_{k\in \mathbb{Z}}\sum_{l\in \mathbb{Z}}\left\langle
C_{k-l,(1,0)}^{X}\left( \Phi _{2k+1,m}^{1}\right) ,\Phi
_{2l,m}^{1}\right\rangle _{\mathcal{H}}  \notag \\
&&+\sum_{k\in \mathbb{Z}}\sum_{l\in \mathbb{Z}}\left\langle
C_{k-l,(0,1)}^{X}\left( \Phi _{2k,m}^{1}\right) ,\Phi
_{2l+1,m}^{1}\right\rangle _{\mathcal{H}}  \notag \\
&&+\sum_{k\in \mathbb{Z}}\sum_{l\in \mathbb{Z}}\left\langle
C_{k-l,(0,0)}^{X}\left( \Phi _{2k+1,m}^{1}\right) ,\Phi
_{2l+1,m}^{1}\right\rangle _{\mathcal{H}},\text{ \ \ \ \ }t\overset{2}{
\equiv }1.\nn
\end{eqnarray}
The long-run covariance matrix is given by
\begin{equation*}
\lim_{n\rightarrow \infty }\frac{1}{n}\mathrm{Var}\left( \mathbf{Y}
_{1}+\cdots +\mathbf{Y}_{n}\right) =\dfrac{2\pi }{2} \left[ \mathrm{diag}
\left( \lambda _{0,1} ,\ldots ,\lambda _{0,p} \right) +\mathrm{\ diag}\left(
\lambda _{0,p+1} ,\ldots ,\lambda _{0,2p} \right) \right] .
\end{equation*}

The following theorem provides a formula for reconstructing the original $
\mathcal{H}$-valued process $X$\ from its DFPC scores $\left\{ Y_{t,m},t\in
\mathbb{Z},m\geq 1\right\} $.

\begin{theorem}
\label{T Inversion Formula} (Inversion Formula) Let $X=\left\{ X_{t},t\in
\mathbb{Z}\right\} $ be an $\mathcal{H}$-valued mean zero $T$-PC
 random process, and $\left\{ Y_{t,m},t\in \mathbb{Z},m\geq 1\right\}
 $ be its DFPC scores. For each time $t$\ and positive integer $m$
 define $X_{t,m}$ to be
\begin{eqnarray*}
X_{t,m}&:&=\sum_{l\in \mathbb{Z}}Y_{t+lT-d,m}\Phi _{lT-d,m}^{0}+\sum_{l\in \mathbb{
Z}}Y_{t+lT-d+1,m}\Phi _{lT-d+1,m}^{1}\\
&&+\cdots+\sum_{l\in \mathbb{Z}}Y_{t+lT-d+T-1,m}\Phi _{lT-d+T-1,m}^{T-1}, \text{\ \ \ \ }  t\overset{T}{\equiv }d
\end{eqnarray*}
Then,
\begin{eqnarray*}
X_{t} &=&\sum_{m\geq 1}X_{t,m} , \text{\ \ \ \ }   t\overset{T}{\equiv }d,
\end{eqnarray*}
where the  convergence holds in mean square provided that
\begin{equation}
\sum_{d=0}^{T-1}\sum_{l\in \mathbb{Z}}\left\Vert \Phi _{l,m}^{d}\right\Vert_{\mathcal{H}}
<\infty .
\end{equation}
\end{theorem}

If $T=2$, then
\begin{equation*}
X_{t,m}:=\sum_{l\in \mathbb{Z}}Y_{t+2l,m}\Phi _{2l,m}^{0}+\sum_{l\in \mathbb{
\ \ Z}}Y_{t+2l+1,m}\Phi _{2l+1,m}^{1}, \text{\ \ \ \ } t\overset{2}{\equiv }0
\end{equation*}
\begin{equation*}
X_{t,m}:=\sum_{l\in \mathbb{\ Z}}Y_{t+2l-1,m}\Phi _{2l-1,m}^{0} +\sum_{l\in
\mathbb{Z}}Y_{t+2l,m}\Phi _{2l,m}^{1}, \text{\ \ \ \ } t\overset{2}{\equiv }
1.
\end{equation*}

The following theorem establishes an optimality property of the above DFPC
filter based on a mean square distance criterion.

\begin{theorem} \label{T Optimality} (Optimality) Let
$X=\left\{ X_{t},t\in \mathbb{Z}\right\}
$ be an $\mathcal{H}$-valued mean zero $T$-PC\ random process,
and $\left\{
X_{t,m},t\in \mathbb{Z},m\geq 1\right\} $ be as in Theorem
\ref{T Inversion Formula}.

For arbitrary $\mathcal{H}$-valued  sequences
\[
\left\{ \Psi _{l,m}^{t},t=0,\ldots,T-1,m\geq
1,l\in
\mathbb{Z}\right\} \ \ {\rm and} \ \
\left\{ \Upsilon _{l,m}^{t},t=0,\ldots,T-1,m\geq 1,l\in
\mathbb{Z}\right\}
\]
with $\sum_{t=0}^{T-1}\sum_{l\in
\mathbb{Z}}\left\Vert \Psi _{l,m}^{t}\right\Vert <\infty $\ and $
\sum_{t=0}^{T-1}\sum_{l\in \mathbb{Z}}\left\Vert \Upsilon
_{l,m}^{t}\right\Vert <\infty $,  for each $m$,  consider
\begin{eqnarray*}
\widetilde{Y}_{t,m} &=&\sum_{l\in \mathbb{Z}}\left\langle X_{\left(
t-l\right) },\Psi _{l,m}^{d}\right\rangle\\
&=&\sum_{l\in \mathbb{Z}}\left\langle X_{\left(
t-lT-d\right) },\Psi _{lT+d,m}^{d}\right\rangle+\sum_{l\in \mathbb{Z}}\left\langle X_{\left(
t-lT-d+1\right) },\Psi _{lT+d-1,m}^{d}\right\rangle\\
&&+\cdots+\sum_{l\in \mathbb{Z}}\left\langle X_{\left(
t-lT-d+T-1\right) },\Psi _{lT+d-T+1,m}^{d}\right\rangle  , \text{\ \ \ \ }  t\overset{T}{\equiv }d
\end{eqnarray*}
and
\begin{eqnarray*}
\widetilde{X}_{t,m}
&=&\sum_{l\in \mathbb{Z}}\widetilde{Y}_{t+lT-d,m}\Upsilon
_{lT-d,m}^{0}+\sum_{l\in \mathbb{Z}}\widetilde{Y}_{t+lT-d+1,m}\Upsilon
_{lT-d+1,m}^{1}\\
&&+\cdots+\sum_{l\in \mathbb{Z}}\widetilde{Y}_{t+lT-d+T-1,m}\Upsilon
_{lT-d+T-1,m}^{T-1} , \text{\ \ \ \ } t\overset{T}{\equiv }d
\end{eqnarray*}

Then,  the following inequality
holds for each $ t \in \mathbb{Z} $ and $p\geq 1$.
\begin{eqnarray*}
&&E\left\Vert X_{Tt}-\sum_{m=1}^{p}X_{Tt,m}\right\Vert ^{2}+\cdots+E\left\Vert
X_{Tt+T-1}-\sum_{m=1}^{p}X_{Tt+T-1,m}\right\Vert ^{2}\\
&=&\sum_{m>p}\int_{-\pi }^{\pi }\lambda _{m}\left( \theta \right) d\theta\\
&\leq& E\left\Vert
X_{Tt}-\sum_{m=1}^{p}\widetilde{X}_{Tt,m}\right\Vert ^{2}+\cdots+E\left\Vert
X_{Tt+T-1}-\sum_{m=1}^{p}\widetilde{X}_{Tt+T-1,m}\right\Vert ^{2}.
\end{eqnarray*}
\end{theorem}

\medskip

In practice, the scores $Y_{t,m}$ are estimated by truncated sums of the
form
\begin{align}
\widehat{Y}_{t,m}& =\sum_{l=-LT+d-T+1}^{LT+d}\left\langle X_{t-l},\widehat{
\Phi }_{l,m}^{d}\right\rangle  \label{Y hat general T} \\
& =\sum_{l=-L}^{L}\left\langle X_{t-lT-d},\widehat{\Phi }_{lT+d,m}^{d}\right
\rangle +\cdots  \notag \\
& \ \ +\sum_{l=-L}^{L}\left\langle X_{t-lT-d+T-1},\widehat{\Phi }
_{lT+d-T+1,m}^{d}\right\rangle ,\text{\ }t\overset{T}{\equiv }d,  \notag
\end{align}
in which $\widehat{\Phi }_{l,m}^{d}$\ s are obtained from an estimator $
\widehat{\mathcal{F}}_{\theta ,\left( q,r\right) }^{X}$. In an asymptotic
setting, the truncation level $L$ is treated as an increasing function of $n$
\ (the length of the time series). (Recommendations for the selection of $L$
in finite samples are discussed in Sections \ref{s:ni} and \ref{s:a}.)

We conclude this section by showing that under mild assumptions, $E\left |
\widehat Y_{t, m} - Y_{t, m}\right | \to 0$, Theorem~\ref{T Consistency}. We
first state the conditions under which the above result holds. Condition~\ref
{Consistency of F hat s} is intuitive, the spectral density operator must be
consistently estimated. The other two conditions are more technical, but
also intuitively clear.

\begin{condition}\label{Consistency of F hat s}
The estimator $\widehat{\mathcal{F}}_{\theta ,\left(
q,r\right) }^{X}$ satisfies
\begin{equation*}
\int_{-\pi }^{\pi }E\left\Vert \mathcal{F}_{\theta ,\left( q,r\right) }^{X}-
\widehat{\mathcal{F}}_{\theta ,\left( q,r\right) }^{X}\right
\Vert _{\mathcal{S}}^{2}\longrightarrow 0,
\text{ \textrm{as }}n\to 0; \text{\ \ \ \ }  q,r=0,\ldots,T-1.
\end{equation*}
\end{condition}

\begin{condition}
\label{distinct eigenvalues} Let $\lambda _{\theta ,m}$
be as in
\eqref{eigenvf decomposition}
and define $\alpha _{\theta ,1}:=\lambda _{\theta ,1}-\lambda _{\theta
,2}$ and $\alpha _{\theta ,m}:=\min \left\{ \lambda _{\theta
,m-1}-\lambda _{\theta ,m},\lambda _{\theta ,m}-\lambda _{\theta
,m+1}\right\} $, $m>1$.  Assume that for each $m$, set $\left\{ \alpha
_{\theta ,m}:\theta \in \left(-\pi,\pi \right]\right \} $ has at most
finitely many zeros.
\end{condition}
Condition \ref{distinct eigenvalues} assures that for almost all $\theta \in
\left(-\pi,\pi \right] $ the $m $-th eigenspace is one--dimensional.

\begin{condition}
\label{suitable orientation} Let $\varphi _{\theta ,m}$ be as defined in
\eqref{eigenvf decomposition}
and $\omega $\ be a given element in $\mathcal{H}^{T}$. Set the orientation
of $\varphi _{\theta ,m}$\ such that $\left\langle \varphi _{\theta
,m},\omega \right\rangle _{\mathcal{H}^{T}}\in \left( 0,\infty \right) $\
whenever $\left\langle \varphi _{\theta ,m},\omega \right\rangle _{\mathcal{H
}^{T}}\neq 0$. Assume $\mathcal{L}eb\left\{ \theta :\left\langle \varphi
_{\theta ,m},\omega \right\rangle _{\mathcal{H}^{T}}=0\right\} =0$,\ where $
\mathcal{L}eb$\ stands for the Lebesgue measure on $\mathbb{R}$\ restricted
to the interval $\left( -\pi ,\pi \right] $.
\end{condition}
Under Condition \ref{suitable orientation} we can set a specific orientation
of eigenfunctions $\varphi_{\theta,m} $ and avoid considering different
versions of DFPC's.

\begin{theorem}\label{T Consistency}
If Conditions
  \ref{Consistency of F hat s}-\ref{suitable orientation}\ hold, then
there is an increasing function $L= L(n)$ such that
$E\left | \widehat Y_{t, m} - Y_{t, m}\right | \to 0$, as $n\to\infty$.
\end{theorem}

\section{Numerical implementation}\label{s:ni}
The theory presented in Section~\ref{s:DFPC} is developed in the
framework of infinite dimensional Hilbert spaces in which the
various functional objects live. Practically usable methodology
requires a number of dimension reduction steps to create
approximating finite dimensional objects which can be manipulated
by computers. This section describes the main steps of such a reduction.
We developed an \verb|R| package, \verb|pcdpca|,  which
allows to preform all procedures described in this paper. In particular,
it is used to perform the analysis and simulations in Section~\ref{s:a}.

We use linearly independent basis functions $\left\{
B_{1},B_{2},\ldots ,B_{K}\right\} $ to convert the data observed at
discrete time points to functional objects of the form $x\left(
u\right) =\sum_{j=1}^{K}c_{j}B_{j}\left( u\right) $. This is just the
usual basis expansion step, see e.g. Chapter 3 of
\citetext{ramsay:hooker:graves:2009} or Chapter 1 of
\citetext{kokoszka:reimherr:2017}. We thus work in a finite
dimensional space $\mathcal{H}_{K}=\mathrm{sp}\left\{
B_{1},B_{2},\ldots ,B_{K}\right\} $. To each bounded linear operator
$A:\mathcal{H} _{K}\rightarrow \mathcal{H}_{K}$ there corresponds  a
complex-valued $ K\times K$\ matrix $\mathfrak{A}$ defined by the
relation $A(x)=\boldsymbol{B }^{\prime }\mathfrak{A}\mathbf{c},$ where
$\boldsymbol{B}=\left( B_{1},B_{2},\ldots ,B_{K}\right) ^{\prime }$
and $\mathbf{c}=\left( c_{1},c_{2},\ldots ,c_{K}\right) ^{\prime
}$. Let $\mathbf{M}_{B}$ be the complex-valued $K\times K$ matrix
$\left( \left\langle B_{q},B_{r}\right\rangle \right) _{q,r=0,\ldots
,K}$, $X_{t}=\boldsymbol{B} ^{\prime }\mathbf{c}_{t}$, and define
\begin{equation*}
\mathbf{B}_{T}^{\prime }:=\left( \left(
\begin{array}{c}
B_{1} \\
0 \\
\vdots \\
0
\end{array}
\right) ,\ldots ,\left(
\begin{array}{c}
B_{K} \\
0 \\
\vdots \\
0
\end{array}
\right) ,\ldots ,\left(
\begin{array}{c}
0 \\
\vdots \\
0 \\
B_{1}
\end{array}
\right) ,\ldots ,\left(
\begin{array}{c}
0 \\
\vdots \\
0 \\
B_{K}
\end{array}
\right) \right) =\left( \mathbf{b}_{1}^{\prime },\mathbf{b}_{2}^{^{\prime
}},\ldots ,\mathbf{b}_{T}^{\prime }\right) .
\end{equation*}
Next, define the matrix
\begin{equation}
\mathfrak{F}_{\theta }^{\underline{X}}=\left(
\begin{array}{ccc}
\mathcal{F}_{\theta ,\left( 0,0\right) }^{\mathbf{c}} & \cdots & \mathcal{F}
_{\theta ,\left( 0,T-1\right) }^{\mathbf{c}} \\
\vdots & \ddots & \vdots \\
\mathcal{F}_{\theta ,\left( T-1,0\right) }^{\mathbf{c}} & \cdots & \mathcal{F
}_{\theta ,\left( T-1,T-1\right) }^{\mathbf{c}}
\end{array}
\right) \left(
\begin{array}{ccc}
\mathbf{M}_{B}^{\prime } &  & 0 \\
& \ddots &  \\
0 &  & \mathbf{M}_{B}^{\prime }
\end{array}
\right)  \label{T matrix of F^X}
\end{equation}
as the matrix corresponding to the operator $\mathcal{F}_{\theta }^{
\underline{X}}$,\ restricted to the subspace $\mathcal{H}_{K}^{T}$. Recall
the definition of the spectral density operators $\mathcal{F}_{\theta ,\left(
q,r\right) }^{\mathbf{c}}$ corresponding to T-PC sequence $\mathbf{c}
=\left\{ \mathbf{c}_{t},t\in \mathbb{Z}\right\} $ from
\eqref{SDO components}.

If $\lambda _{\theta ,m}$\ and $\mathbf{\varphi }_{\theta ,m}:=\left(
\mathbf{\varphi }_{\theta ,m,1}^{\prime },\ldots ,\mathbf{\ \varphi }
_{\theta ,m,T}^{\prime }\right) ^{\prime }$\ are the $m$-th
eigenvalue and
eigenfunction of $TK\times TK$\ complex-valued
matrix
$\mathfrak{F}_{\theta
}^{\underline{X}}$, then $\lambda _{\theta ,m}$ and
\[\mathbf{B}_{T}^{\prime
}\mathbf{\varphi }_{\theta ,m}=\left( \mathbf{b}_{1}^{\prime },\mathbf{b}
_{2}^{\prime },\ldots ,\mathbf{b}_{T}^{\prime }\right) \left( \mathbf{
\varphi }_{\theta ,m,1}^\prime,\ldots ,\mathbf{\varphi }_{\theta
,m,T}^{\prime }\right) ^\prime=\left( \boldsymbol{B}^{\prime }\mathbf{
\varphi }_{\theta ,m,1},\ldots ,\boldsymbol{B}^{\prime }\mathbf{\varphi }
_{\theta ,m,T}\right)^\prime
\]
are the $m$-th eigenvalue and
eigenfunction of the operator $\mathcal{F}_{\theta }^{\underline{X}}$. This
motivates us to use the ordinary multivariate techniques to estimate $
C_{h,\left( q,r\right) }^{\mathbf{c}}$ and consequently $\mathcal{F}
_{\theta ,\left( q,r\right) }^{\mathbf{c}}$, for $q,r=0,\ldots ,T-1$, and $
\theta \in \left( -\pi ,\pi \right] $, by
\[
\widehat{C}_{h,\left( q,r\right) }^{\mathbf{c}}=\frac{T}{n}\sum_{j\in
\mathbb{Z}}\mathbf{c}_{q+Tj}\mathbf{c}_{r+Tj-Th}^{\prime }I\left\{ 1\leq
q+Tj\leq n\right\} I\left\{ 1\leq r+Tj-Th\leq n\right\} ,\text{ \ \ \ }h\geq
0,
\]
\begin{equation*}
\left( \overline{\widehat{C}}_{-h,\left( q,r\right) }^{\mathbf{c}}\right)
^{\prime }=\widehat{C}_{h,\left( q,r\right) }^{\mathbf{c}},
\text{ \ \ \ }h<0,
\end{equation*}
and
\begin{equation}
\widehat{\mathcal{F}}_{\theta ,\left( q,r\right) }^{\mathbf{c}}=\frac{1}{
2\pi }\sum_{\left\vert h\right\vert \leq q(n)}w\left( \frac{h}{q(n)}\right)
\widehat{C}_{h,\left( q,r\right) }^{\mathbf{c}}e^{-ih\theta },
\label{T F hat of C}
\end{equation}
where $w$ is a suitable weight function, $q(n)\longrightarrow \infty $, and $
\frac{q(n)}{n}\longrightarrow 0$. Replace \eqref{T F hat of C} by
\eqref{T
matrix of F^X} to obtain a\textbf{\ }consistent estimator $\widehat{
\mathfrak{F}}_{\theta }^{\underline{X}}$ with eigenvalues $\widehat{\lambda }
_{\theta ,m}$ \ and eigenvectors $\widehat{\mathbf{\varphi }}_{\theta
,m}:=\left( \widehat{\mathbf{\varphi }}_{\theta ,m,1}^{\prime },\ldots ,
\widehat{\mathbf{\ \varphi }}_{\theta ,m,T}^{\prime }\right) ^{\prime }$, $
m\geq 1$. Use
\begin{eqnarray*}
\mathbf{B}_{T}^{\prime }\widehat{\mathbf{\varphi }}_{\theta ,m} &=&\left(
\mathbf{b}_{1}^{\prime },\ldots ,\mathbf{b}_{T}^{\prime }\right) \left(
\begin{array}{c}
\widehat{\mathbf{\varphi }}_{\theta ,m,1} \\
\vdots \\
\widehat{\mathbf{\varphi }}_{\theta ,m,T}
\end{array}
\right)
=\left(
\begin{array}{c}
\boldsymbol{B}^{\prime }\widehat{\mathbf{\varphi }}_{\theta ,m,1} \\
\vdots \\
\boldsymbol{B}^{\prime }\widehat{\mathbf{\varphi }}_{\theta ,m,T}
\end{array}
\right)
=\left(
\begin{array}{c}
\widehat{\varphi }_{\theta ,m,1} \\
\vdots \\
\widehat{\varphi }_{\theta ,m,T}
\end{array}
\right) =\widehat{\varphi }_{\theta ,m}
\end{eqnarray*}
to get estimators $\widehat{\varphi }_{\theta ,m}$,  and set
\begin{equation*}
\left(
\begin{array}{c}
\boldsymbol{B}^{\prime }\widehat{\mathbf{\varphi }}_{\theta ,dp+m,1} \\
\vdots \\
\boldsymbol{B}^{\prime }\widehat{\mathbf{\varphi }}_{\theta ,dp+m,T}
\end{array}
\right) =\left(
\begin{array}{c}
\widehat{\Phi }_{\theta ,d,m}^{d} \\
\vdots \\
\widehat{\Phi }_{\theta ,d-T+1,m}^{d}
\end{array}
\right),
\end{equation*}
or equivalently
\begin{equation}
\left(
\begin{array}{c}
\boldsymbol{B}^{\prime }\widehat{\mathbf{\Phi }}_{lT+d,m}^{d} \\
\vdots \\
\boldsymbol{B}^{\prime }\widehat{\mathbf{\Phi }}_{lT+d-T+1,m}^{d}
\end{array}
\right) :=\dfrac{1}{2\pi }\int_{-\pi }^{\pi }\left(
\begin{array}{c}
\boldsymbol{B}^{\prime }\widehat{\mathbf{\varphi }}_{\theta ,dp+m,1} \\
\vdots \\
\boldsymbol{B}^{\prime }\widehat{\mathbf{\varphi }}_{\theta ,dp+m,T}
\end{array}
\right) e^{-il\theta }d\theta =\left(
\begin{array}{c}
\widehat{\Phi }_{lT+d,m}^{d} \\
\vdots \\
\widehat{\Phi }_{lT+d-T+1,m}^{d}
\end{array}
\right) ,  \label{T Phi hats}
\end{equation}
for $d=0,\ldots ,T-1$, $m=1,\ldots ,p$. Note that one may use numerical
integration to find $\widehat{\Phi }_{l,m}^{d}$. Therefore, the PC-DFPC
scores can be estimated by
\begin{eqnarray*}
\widehat{Y}_{t,m} &=&\sum_{l=-LT+d-T+1}^{LT+d}\left\langle X_{t-l},\widehat{
\Phi }_{l,m}^{d}\right\rangle\\
&=&\sum_{l=-L}^{L}\left\langle X_{t-lT-d},\widehat{\Phi }_{lT+d,m}^{d}\right
\rangle +\cdots +\sum_{l=-L}^{L}\left\langle X_{t-lT-d+T-1},\widehat{\Phi }
_{lT+d-T+1,m}^{d}\right\rangle  \notag \\
&=&\sum_{l=-L}^{L}\mathbf{c}_{\left( t-lT-d\right) }^{\prime }\mathbf{M}_{B}
\overline{\widehat{\mathbf{\Phi }}}_{lT+d,m}^{d}+\ldots +\sum_{l=-L}^{L}
\mathbf{c}_{\left( t-lT-d+T-1\right) }^{\prime }\mathbf{M}_{B}\overline{
\widehat{\mathbf{\Phi }}}_{lT+d-T+1,m}^{d},\text{\ }t\overset{T}{\equiv }d,
\end{eqnarray*}
where $L$ satisfies
\begin{equation*}
\sum_{l=-L}^{L}\left( \left\Vert \widehat{\Phi }_{lT+d,m}^{d}\right\Vert _{
\mathcal{H}}^{2}+\cdots +\left\Vert \widehat{\Phi }_{lT+d-T+1,m}^{d}\right
\Vert _{\mathcal{H}}^{2}\right) \geq 1-\varepsilon
\end{equation*}
\ for some\textbf{\ }$d$ and small $\varepsilon >0$. Consequently, $X_{t}$
can be approximated by
\begin{equation*}
\widehat{X}_{t}=\sum_{m=1}^{p}\sum_{l=-L}^{L}\widehat{Y}_{t+lT-d,m}\widehat{
\Phi }_{lT-d,m}^{0}+\ldots +\sum_{m=1}^{p}\sum_{l=-L}^{L}\widehat{Y}
_{t+lT-d+T-1,m}\widehat{\Phi }_{lT-d+T-1,m}^{T-1},\text{ \ \ \ }t\overset{T}{
\equiv }d.
\end{equation*}
\begin{remark}
Usually, the mean $\mu $ of the process $X$\ is not known. In this case we
first use smoothed functions $X_{t}=\boldsymbol{B}^\prime
\mathbf{c}_{t}$ to obtain estimators $\widehat{\mu }_{0},\ldots,\widehat{\mu }_{T-1}$
or $T$-periodic mean function estimator $\left\{ \widehat{\mu
}_{t}:\widehat{\mu }_{Tk+d}=\widehat{\mu }_{d};k\in
\mathbb{Z}\right\} $, then apply the above method to the centered functional
observations $X_{t}^{\ast }=X_{t}-\widehat{\mu }_{t}$.
\end{remark}

For  illustration,  set $T=2$ and define
\begin{equation*}
\mathbf{B}_{T}^{\prime }=\left( \left(
\begin{array}{c}
B_{1} \\
0
\end{array}
\right) ,\ldots ,\left(
\begin{array}{c}
B_{K} \\
0
\end{array}
\right) ,\left(
\begin{array}{c}
0 \\
B_{1}
\end{array}
\right) ,\ldots ,\left(
\begin{array}{c}
0 \\
B_{K}
\end{array}
\right) \right) =\left( \mathbf{b}_{1}^{\prime },\mathbf{b}_{2}^{^{\prime
}}\right)
\end{equation*}
as a vector of linearly independent elements in\ $\mathcal{H}^{2}$.
The matrix
\begin{equation} \label{matrix of F^X}
\mathfrak{F}_{\theta }^{\underline{X}}=\left(
\begin{array}{cc}
\mathcal{F}_{\theta ,\left( 0,0\right) }^{\mathbf{c}} & \mathcal{\ F}
_{\theta ,\left( 0,1\right) }^{\mathbf{c}} \\
\mathcal{F}_{\theta ,\left( 1,0\right) }^{\mathbf{c}} & \mathcal{\ F}
_{\theta ,\left( 1,1\right) }^{\mathbf{c}}
\end{array}
\right) \left(
\begin{array}{cc}
\mathbf{M}_{B}^{\prime } & 0 \\
0 & \mathbf{M}_{B}^{\prime }
\end{array}
\right)
\end{equation}
corresponds  to the operator $\mathcal{F}_{\theta }^{
\underline{X}}$\ restricted to to the subspace $\mathcal{H}_{K}^{2}$.
For $q,r=0,1$,
\[
\widehat{C}_{h,\left( q,r\right) }^{\mathbf{c}}=\frac{2}{n}\sum_{j\in
\mathbb{Z}}\mathbf{c}_{q+2j}\mathbf{c}_{r+2j-2h}^{\prime }I\left\{ 1\leq
q+2j\leq n\right\} I\left\{ 1\leq r+2j-2h\leq n\right\} ,\text{ \ \ \ }h\geq
0,
\]
Estimators of  the PC-DFPC filter coefficient $\Phi _{l,m}^{d}$
and the PC-DFPC scores $Y_{t,m}$ are
\begin{eqnarray*}
\left(
\begin{array}{c}
\boldsymbol{B}^{\prime }\widehat{\mathbf{\Phi }}_{2l,m}^{0} \\
\boldsymbol{B}^{\prime }\widehat{\mathbf{\Phi }}_{2l-1,m}^{0}
\end{array}
\right) &:&=\dfrac{1}{2\pi }\int_{-\pi }^{\pi }\left(
\begin{array}{c}
\boldsymbol{B}^{\prime }\widehat{\mathbf{\varphi }}_{\theta ,m,1} \\
\boldsymbol{B}^{\prime }\widehat{\mathbf{\varphi }}_{\theta ,m,2}
\end{array}
\right) e^{-il\theta }d\theta =\left(
\begin{array}{c}
\widehat{\Phi }_{2l,m}^{0} \\
\widehat{\Phi }_{2l-1,m}^{0}
\end{array}
\right) \\
\left(
\begin{array}{c}
\boldsymbol{B}^{\prime }\widehat{\mathbf{\Phi }}_{2l,m+p}^{1} \\
\boldsymbol{B}\widehat{\mathbf{\Phi }}_{2l-1,m+p}^{1}
\end{array}
\right) &:&=\dfrac{1}{2\pi }\int_{-\pi }^{\pi }\left(
\begin{array}{c}
\boldsymbol{B}^{\prime }\widehat{\mathbf{\varphi }}_{\theta ,m+p,1} \\
\boldsymbol{B}^{\prime }\widehat{\mathbf{\varphi }}_{\theta ,m+p,2}
\end{array}
\right) e^{-il\theta }d\theta =\left(
\begin{array}{c}
\widehat{\Phi }_{2l+1,m}^{1} \\
\widehat{\Phi }_{2l,m}^{1}
\end{array}
\right) ,\text{ \ \ \ }m=1,\ldots ,p.
\end{eqnarray*}
\begin{eqnarray*}
\widehat{Y}_{t,m} &=&\sum_{l=-2L-1}^{2L}\left\langle X_{\left( t-l\right) },
\widehat{\Phi }_{l,m}^{0}\right\rangle \\
&=&\sum_{l=-L}^{L}\mathbf{c}_{\left( t-2l\right) }^{\prime }\mathbf{M}_{B}
\overline{\widehat{\mathbf{\Phi }}}_{2l,m}^{0}+\sum_{l=-L}^{L}\mathbf{c}
_{\left( t-2l+1\right) }^{\prime }\mathbf{M}_{B}\overline{\widehat{\mathbf{
\Phi }}}_{2l-1,m}^{0},\text{ \ \ \ }t\overset{2}{\equiv }0,
\end{eqnarray*}
\begin{eqnarray*}
\widehat{Y}_{t,m} &=&\sum_{l=-2L}^{2L+1}\left\langle X_{\left( t-l\right) },
\widehat{\Phi }_{l,m}^{1}\right\rangle \\
&=&\sum_{l=-L}^{L}\mathbf{c}_{\left( t-2l\right) }^{\prime }\mathbf{M}_{B}
\overline{\widehat{\mathbf{\Phi }}}_{2l,m}^{1}+\sum_{l=-L}^{L}\mathbf{c}
_{\left( t-2l-1\right) }^{\prime }\mathbf{M}_{B}\overline{\widehat{\mathbf{
\Phi }}}_{2l+1,m}^{1},\text{ \ \ \ }t\overset{2}{\equiv }1.
\end{eqnarray*}
Hence,
\begin{equation*}
\widehat{X}_{t}=\sum_{m=1}^{p}\sum_{l=-L}^{L}\widehat{Y}_{t+2l,m}\widehat{
\Phi }_{2l,m}^{0}+\sum_{m=1}^{p}\sum_{l=-L}^{L}\widehat{Y}_{t+2l+1,m}
\widehat{\Phi }_{2l+1,m}^{1},\text{ \ \ \ }t\overset{2}{\equiv }0,
\end{equation*}
\begin{equation*}
\widehat{X}_{t}=\sum_{m=1}^{p}\sum_{l=-L}^{L}\widehat{Y}_{t+2l,m}\widehat{
\Phi }_{2l,m}^{1}+\sum_{m=1}^{p}\sum_{l=-L}^{L}\widehat{Y}_{t+2l-1,m}
\widehat{\Phi }_{2l-1,m}^{0},\text{ \ \ \ }t\overset{2}{\equiv }1.
\end{equation*}

\section{Application to particulate pollution data and a  simulation
study} \label{s:a}
To illustrate the advantages of PC-DFPCA relative the (stationary)
DFPCA which may arise in certain settings, we further explore the
dataset analyzed in \citetext{hormann:kidzinski:hallin:2015}.  The
dataset contains intraday measurements of pollution in Graz, Austria
between October 1, 2010 and March 31, 2011. Observations were sampled
every $30$ minutes and measure concentration of particle matter of
diameter of less than $10\mu m$ in the ambient air. In order to
facilitate the comparison with the results reported in
\citetext{hormann:kidzinski:hallin:2015}, we employ exactly the same
preprocessing procedure, including square-root transformation, removal
of the mean weakly pattern and outliers. Note that the removal of the
mean weakly pattern does not affect periodic covariances between
weekdays and therefore they can be exploited using the PC-DFPCA
procedure applied with the period $T=7$.  The preprocessed dataset
contains $175$ daily curves, each  converted to a functional object with
$15$ Fourier basis functions, yielding a functional time series $\{X_t
: 1 \leq t \leq 175\}$.

For FPCA and DFPCA we use the same procedure as
\citetext{hormann:kidzinski:hallin:2015} using the implementation
published by those researchers as the \verb|R| package \verb|freqdom|.
To implement the PC-DFPCA, some modifications are needed.  Regarding
the metaparameters $q$ and $L$,
\citetext{hormann:kidzinski:hallin:2015} advise choosing $q \sim
\sqrt{n}$ and $L$ such that $\sum_{-L \leq i \leq L} \| \Phi^d_{m,l}
\|^2 \sim 1$. By design, our estimators of covariances use only around
$n/T$ of all observations and therefore we scale $q$ further by
$\sqrt{T}$, obtaining $q=4$. To avoid overfitting we set $L = 3$
(compared to $10$ in \citetext{hormann:kidzinski:hallin:2015}), as it
now relates to weeks not days and we do not expect dependence beyond
$3$ weeks.

As a measure of fit, we use the  Normalized Mean Squared Error
(\verb|NMSE|) defined as
\[
\verb|NMSE| = \sum_{t=1}^n \|X_t - \hat{X_t}\|^2 /
\sum_{t=1}^n \|X_t\|^2,
\]
where the $\hat{X_t}$ are the observations obtained from the inverse
transform. We refer to the value $\verb|NMSE|\cdot 100\% $ as the {\it
  percentage of variance explained}.

\begin{figure}[!ht]
\centering
\includegraphics[width=0.9\textwidth]{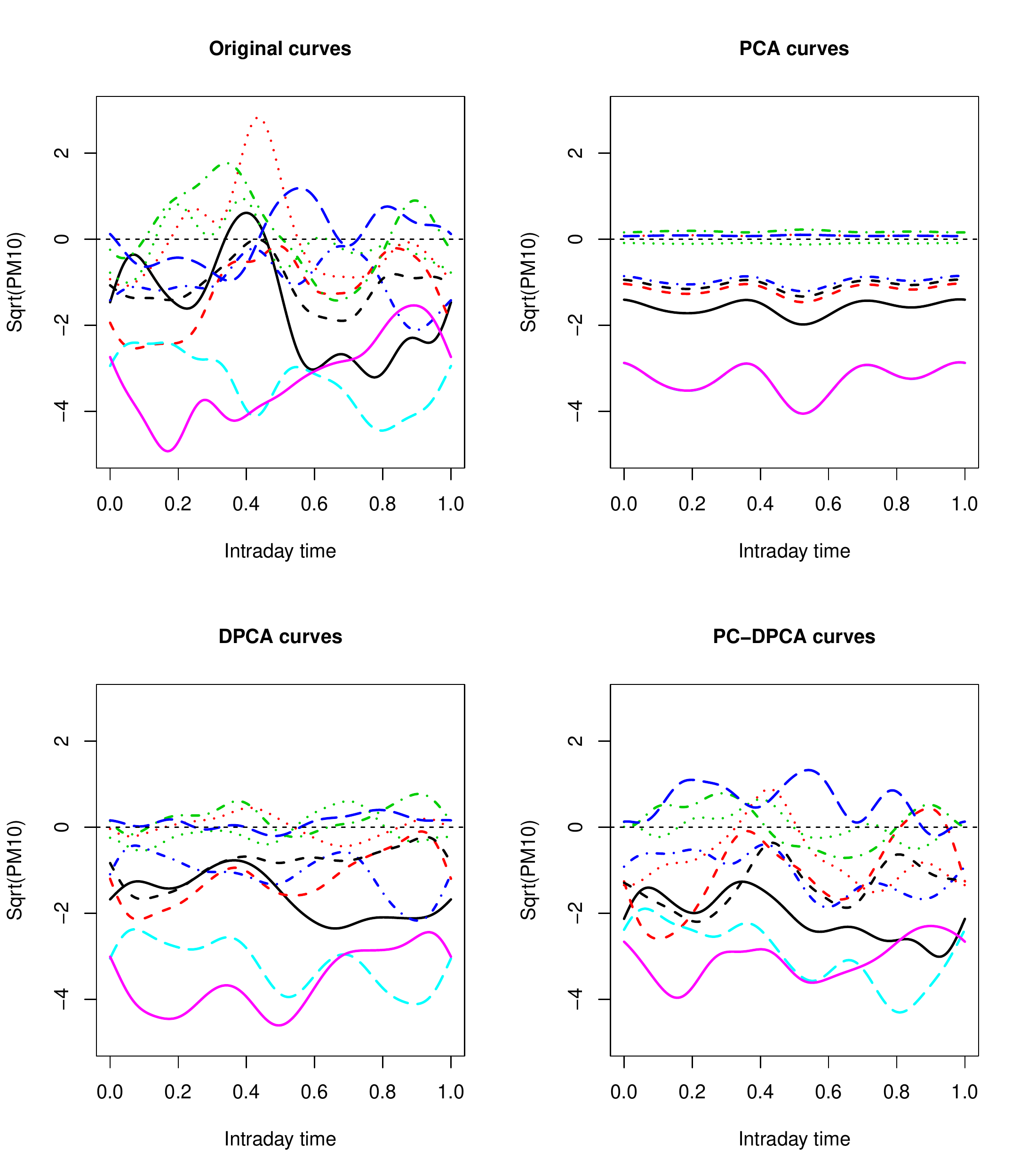}
\caption{\label{fig:curves} Ten successive intraday observations of
  PM10 data (top-left), the corresponding functional PCA curves
  reconstructed  from the first principal component (top-right), dynamic
  functional principal component curves (bottom-left) and
  periodically-correlated dynamic principal components (bottom-right).
  Colors and types of curves match the same observations among plots.}
\end{figure}

For the sake of comparison and discussion,  we focus only on the first
principal component, which already explains $73\%$ of variability in
the static FPCA, $80\%$ of variability in the DFPCA and $88\%$ of
variability in the PC-DFPCA procedure.  Curves corresponding to the
components obtained through each of these methods are presented in
Figure \ref{fig:curves}. As the percentages above suggest, there
is a clear progression in the quality of the approximation using
just one component. This is an important finding because the
purpose of the principal component analysis of any type is
to reduce the dimension of the data using the smallest possible number
of projections without sacrificing the quality of approximation.

\begin{figure}[h!]
\centering
\includegraphics[width=\textwidth]{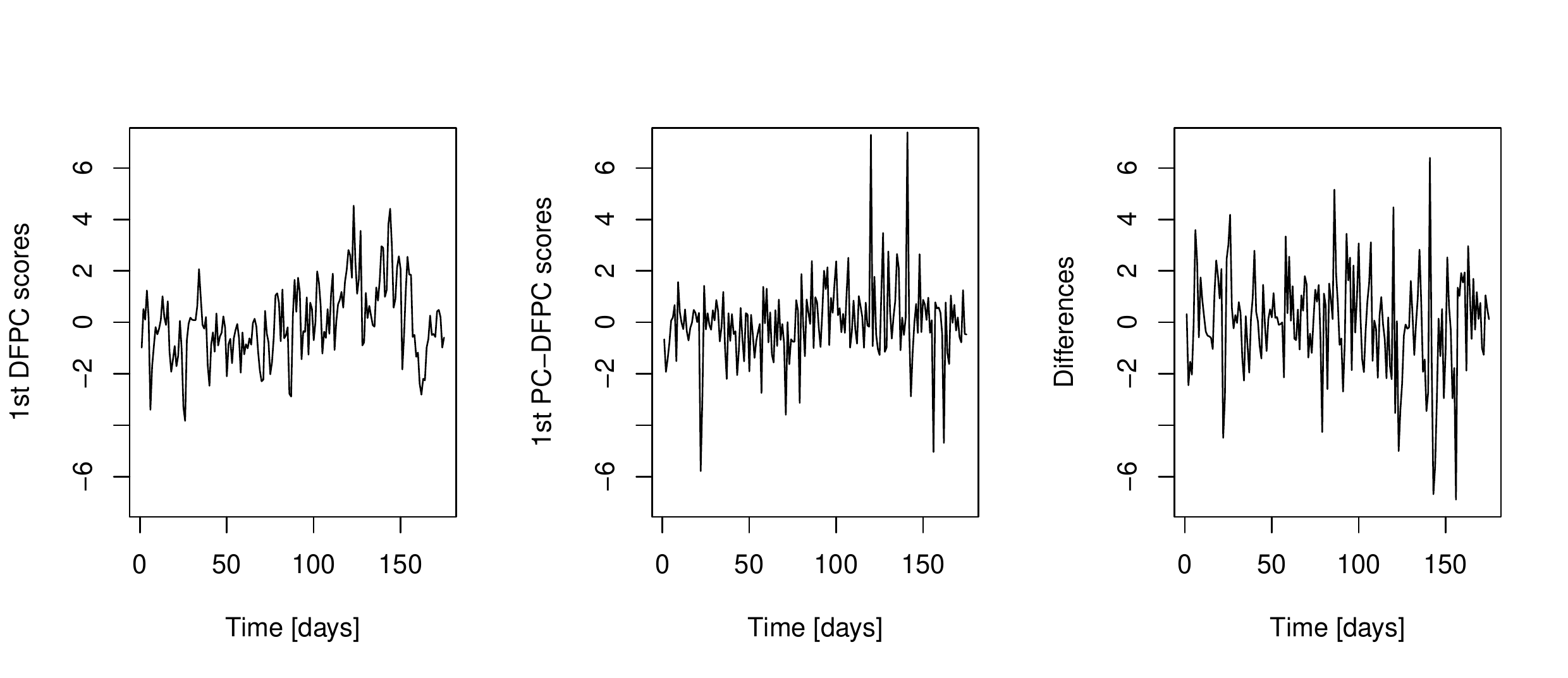}
\caption{\label{fig:scores} The first dynamic principal component scores (left), the first periodically correlated dynamic principal component scores (middle) and differences between the two series (right).}
\end{figure}

\citetext{hormann:kidzinski:hallin:2015} observed that,  for this
particular dataset,  the sequences of scores of the DFPC's
and the static FPC's were almost
identical. This is no longer the case if the PC-DFPC's are used.
Figure \ref{fig:scores} compares the  DFPC and the PC-DFPC scores  and
shows that the resulting time series are quite different. The  PC-DFPCA
takes into account the periodic correlation structure which neither
the static nor the (stationary) DFPCA do.

\begin{figure}[h!]
\centering
\includegraphics[width=\textwidth]{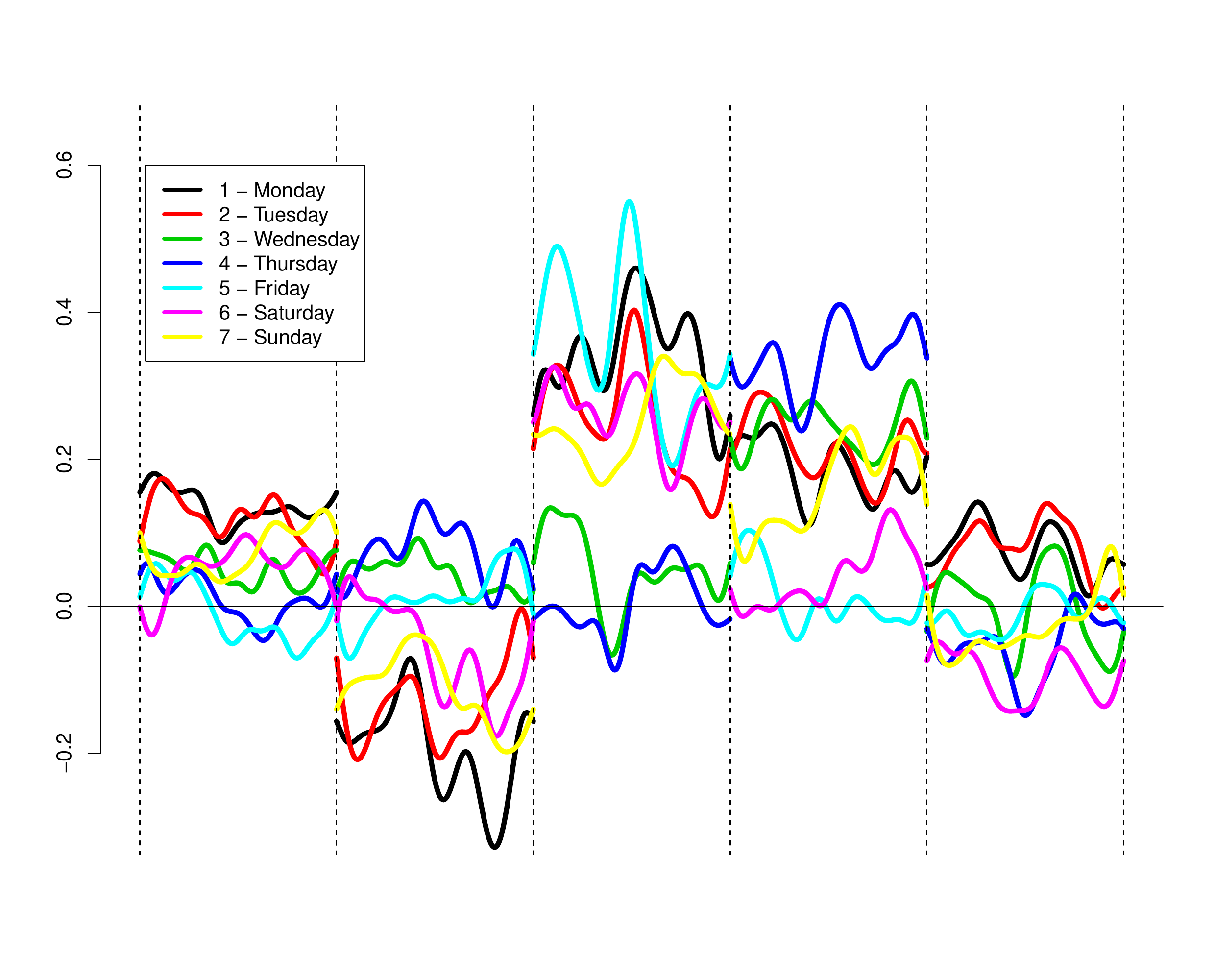}
\caption{\label{fig:filters} Filters of the first principal component
  for $d=0$, i.e. corresponding to Monday. For every $m = 1,2,...,7$
  (as in the legend in parethesis), the five curves of the same color
  correspond to $\Phi^d_{-2,m}, \Phi^d_{-1,m}, ..., \Phi^d_{2,m}$}.
\end{figure}

The estimated PC-DFPCA filters are very high dimensional as can be seen in
Figure \ref{fig:filters}. In particular, with $L = 3$, $T=7$ and $p = 15$
we estimated $(2L~+~1)T^2p^2~=~735$ real numbers,
which may raise  concerns
about  overfitting. This however does not translate into problems
with the finite sample performance, as the following simulation
study shows.

\begin{figure}[h!]
\centering
\includegraphics[width=0.45\textwidth]{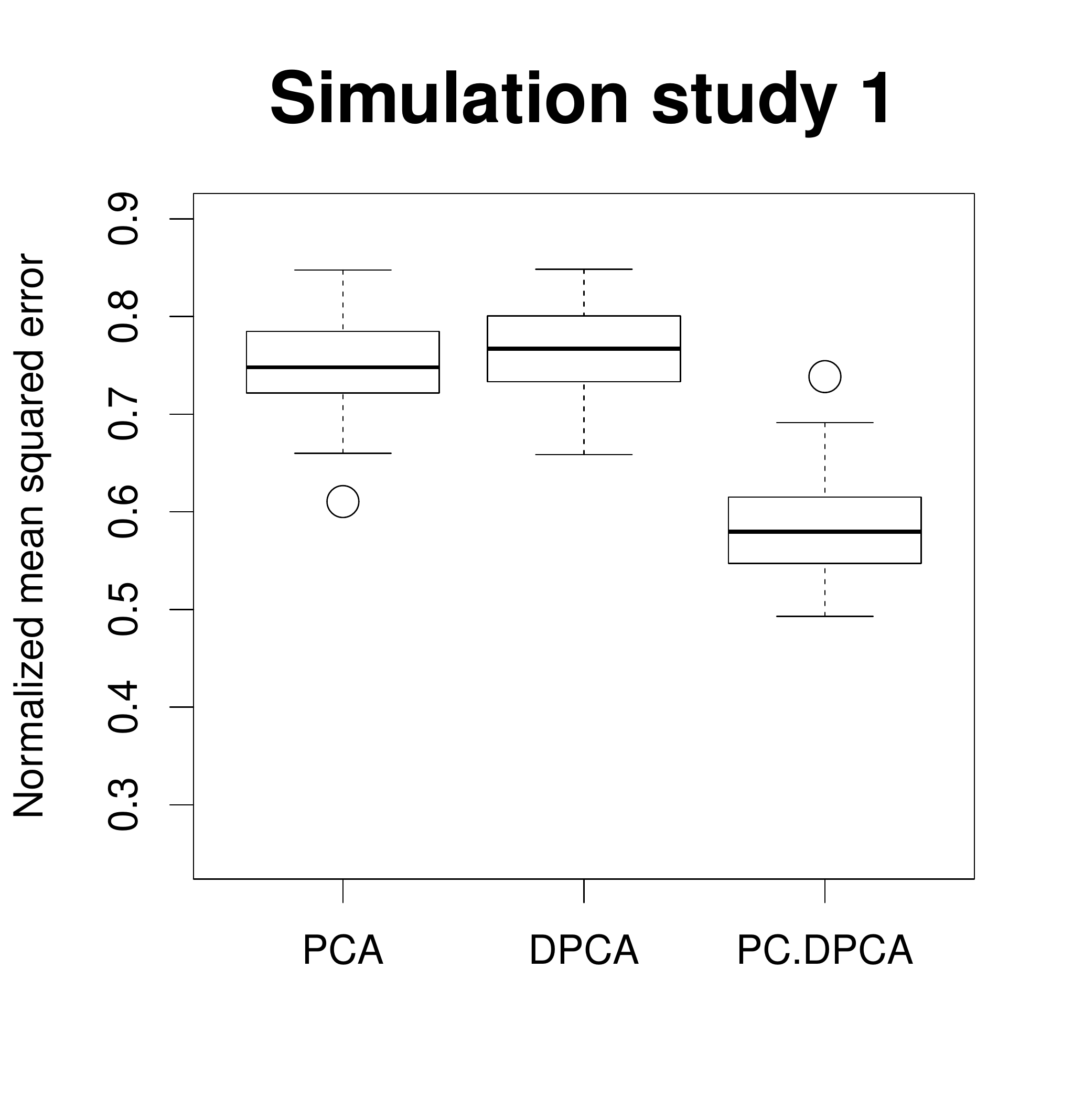}
\includegraphics[width=0.45\textwidth]{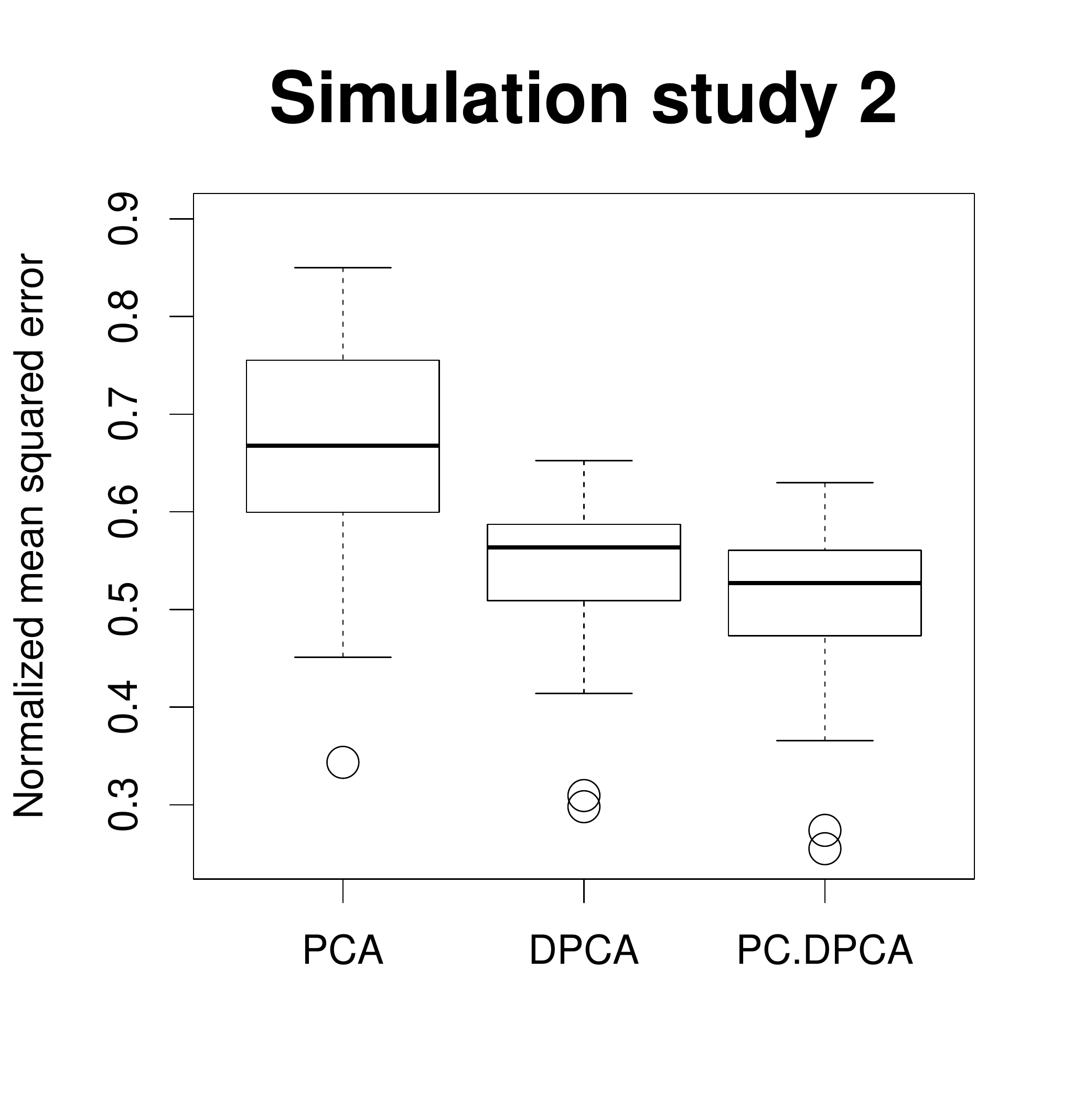}
\caption{\label{fig:simulation}Results of two simulation studies. We
  repeat every simulation 100 times and report distribution of NMSE
  from these repetitions. PC-DFPCA outperforms the two other methods
  in both setups. }
\end{figure}

To further analyze the properties of the PC-DFPCA filter, we design
two simulation studies, with two distinct periodically correlated
functional time series. In the first study we set $p=7$, $T=3$,
$n=300$ and generate variables $\mathbf{a}_i,\mathbf{b}_i \sim
\mathcal{N}(0,\diag(\exp(-2\cdot\frac{1}{p}),\exp(-2\cdot\frac{2}{p}),...,\exp(-2\cdot\frac{p}{p})))$
for $i~\in~\{1,2,...,100\}$. The exponential decay emulates the decay
of typical functional data observations in Fourier basis
representation, where high frequencies tend to $0$. Next, we define
the mulitvariate time series as follows
\[ \mathbf{c}_{Ti + 1} = \mathbf{a}_i, \mathbf{c}_{Ti + 2} =
\mathbf{b}_i \text{ and } \mathbf{c}_{Ti + 3} = 2\mathbf{a}_i -
\mathbf{b}_i \text{ for } i \in \{1,2,...,100\}. \] We divide the set
of $\{\mathbf{c}_t : t \in \{1,2,...,300\}\}$ into a training set
(first $150$ observations) and test set (last $150$ observations). We
train the three methods: FPCA, DFPCA and PC-DFPCA on the training set
and compare the normalized mean squared errors on the test set. We
choose parameters: $L = 2$, $q=3$ and Bartlett weights.  We repeat the
simulation $100$ times and record average NMSE from all runs. As
illustrated in Figure \ref{fig:simulation}, PC-DFPCA outperforms the
two other methods with mean $0.59\ (\sd = 0.05)$ compared to DFPCA
$0.76\ (\sd = 0.04)$ and FPCA $0.74\ (\sd = 0.04)$.

In the second simulation scenario,  we sampled a functional periodically
correlated autoregressive process.  We set $p=7$, $T=2$ and $n=1000$
and generate iid innovations
\[\varepsilon_i \sim \mathcal{N}(0,\Sigma)\]
with
\[
\Sigma = \diag(\exp(-2\cdot1/p,\exp(-2\cdot2/p),...,\exp(-2\cdot p/p)))
\]
for $i \in \{1,2,...,n\}$. We sample 4 operators
$P_{i,j} = [\delta_{k,l}]_{1\leq k,l \leq p}$ and define
$\Psi_{i,j} = 0.9 \cdot P_{i,j} / \|P_{i,j}\|$ for $i,j \in \{0,1\}$,
where $\delta_{k,l} \sim \mathcal{N}(0,\exp(-2\cdot\frac{l}{p}))$.
We define the process $\mathbf{c}_t$ as
\[ \mathbf{c}_t = [0,0,...,0]' \text{ for } t \leq 0 \]
and for $d \in \{0,1\}$ and any $1 \leq t \leq n$ we set
\[ \mathbf{c}_t = \Psi_{d,0}\mathbf{c}_{t-1} +
\Psi_{d,1}\mathbf{c}_{t-2} + \varepsilon_t \text{ for } t
\stackrel{2}{\equiv} d. \] As in the first simulation, we repeat the
experiment $100$ times, record average NMSE and present results in
Figure \ref{fig:simulation}. Again PC-DFPCA outperforms the two other
methods with mean $0.51\ (\sd = 0.07)$ compared to DFPCA $0.55\ (\sd =
0.07)$ and FPCA $0.67\ (\sd = 0.1)$.

\clearpage

\section{Proofs of the results of Section~\ref{s:DFPC}}\label{s:p}
To explain the essence and technique of the proofs, we
consider the special case of $T=2$. The arguments for general $T$ proceed
analogously, merely with a more heavy and less explicit notation, which may
obscure the essential arguments.

\medskip

\noindent \textsc{Proof of Theorem \ref{T filter process}}: To
establish  the mean
square convergence of the series $\sum_{l\in \mathbb{Z}}\Psi _{l}^{t}\left(
X_{\left( t-l\right) }\right) $, let $S_{n,t}$ be its partial sum, i.e. $
S_{n,t}=\sum_{-n\leq l\leq n}\Psi _{l}^{t}\left( X_{\left( t-l\right)
}\right) $, for each  positive integer $n$. Then for $m<n$ we have,
\begin{eqnarray}
E\left\Vert S_{n,t}-S_{m,t}\right\Vert _{\mathbb{C}^{p}}^{2}
&=&\sum_{m<\left\vert l\right\vert ,\left\vert k\right\vert \leq
n}E\left\langle \Psi _{l}^{t}\left( X_{\left( t-l\right) }\right) ,\Psi
_{k}^{t}\left( X_{\left( t-k\right) }\right) \right\rangle _{\mathbb{C}^{p}}
\notag \\
&\leq &\sum_{m<\left\vert l\right\vert ,\left\vert k\right\vert \leq
n}E\left( \left\Vert \Psi _{l}^{t}\left( X_{\left( t-l\right) }\right)
\right\Vert _{\mathbb{C}^{p}}\left\Vert \Psi _{k}^{t}\left( X_{\left(
t-k\right) }\right) \right\Vert _{\mathbb{C}^{p}}\right)  \notag \\
&\leq &\sum_{m<\left\vert l\right\vert ,\left\vert k\right\vert \leq
n}\left\Vert \Psi _{l}^{t}\right\Vert _{\mathcal{L}}\left\Vert \Psi
_{k}^{t}\right\Vert _{\mathcal{L}}E\left( \left\Vert X_{\left( t-l\right)
}\right\Vert \left\Vert X_{\left( t-k\right) }\right\Vert \right)  \notag \\
&\leq &\sum_{m<\left\vert l\right\vert ,\left\vert k\right\vert \leq
n}\left\Vert \Psi _{l}^{t}\right\Vert _{\mathcal{L}}\left\Vert \Psi
_{k}^{t}\right\Vert _{\mathcal{L}}\left( E\left\Vert X_{\left( t-l\right)
}\right\Vert ^{2}E\left\Vert X_{\left( t-k\right) }\right\Vert ^{2}\right) ^{
\frac{1}{2}}  \notag \\
&\leq &M\sum_{\left\vert l\right\vert >m}\sum_{\left\vert k\right\vert
>m}\left\Vert \Psi _{l}^{t}\right\Vert _{\mathcal{L}}\left\Vert \Psi
_{k}^{t}\right\Vert _{\mathcal{L}}\text{ \ \ \ }\mathrm{for}\text{ }\mathrm{
some}\text{ }M\in \mathbb{R}^{+}  \notag \\
&\leq &M\left( \sum_{\left\vert l\right\vert >m}\left\Vert \Psi
_{l}^{t}\right\Vert _{\mathcal{L}}\right) ^{2}.  \label{Tail}
\end{eqnarray}
Summability condition \eqref{PC filter coef} implies that \eqref{Tail}
tends to zero, as $n$ and $m$ tend to infinity.
Therefore, $\left\{S_{n,t},n\in
\mathbb{Z}^{+}\right\} $ forms a Cauchy sequence in
$L^{2}\left( \mathbb{C}
^{p},\Omega \right) $, for each $t$, which implies the desired mean square
convergence. According to the representation of the
filtered process $\mathbf{Y}$ at time $t$\ \textit{i.e.},
\begin{eqnarray*}
\mathbf{Y}_{t} &=&\sum_{l\in \mathbb{Z}}\Psi _{l}^{0}\left( X_{\left(
t-l\right) }\right) \\
&=&\sum_{l\in \mathbb{Z}}\Psi _{2l}^{0}\left( X_{\left( t-2l\right) }\right)
+\sum_{l\in \mathbb{Z}}\Psi _{2l-1}^{0}\left( X_{\left( t-2l+1\right)
}\right) ,\text{\ \ \ \ }t\overset{2}{\equiv }0
\end{eqnarray*}
\begin{eqnarray*}
\mathbf{Y}_{t} &=&\sum_{l\in \mathbb{Z}}\Psi _{l}^{1}\left( X_{\left(
t-l\right) }\right) \\
&=&\sum_{l\in \mathbb{Z}}\Psi _{2l}^{1}\left( X_{\left( t-2l\right) }\right)
+\sum_{l\in \mathbb{Z}}\Psi _{2l+1}^{1}\left( X_{\left( t-2l-1\right)
}\right) ,\text{\ \ \ \ }t\overset{2}{\equiv }1,
\end{eqnarray*}
for each $h\in \mathbb{Z}$ we have
\begin{eqnarray*}
\mathrm{Cov}\left( \mathbf{Y}_{2h},\mathbf{Y}_{0}\right)
&=&\lim_{n\rightarrow \infty }\sum_{\left\vert k\right\vert \leq
n}\sum_{\left\vert l\right\vert \leq n}\mathrm{Cov}\left( \Psi
_{k}^{0}\left( X_{\left( 2h-k\right) }\right) ,\Psi _{l}^{0}\left( X_{\left(
0-l\right) }\right) \right) \\
&=&\sum_{k\in \mathbb{Z}}\sum_{l\in \mathbb{Z}}\Psi _{2k}^{0}\mathrm{Cov}
\left( X_{\left( 2h-2k\right) },X_{-2l}\right) \left( \Psi _{2l}^{0}\right)
^{\ast } \\
&&+\sum_{k\in \mathbb{Z}}\sum_{l\in \mathbb{Z}}\Psi _{2k}^{0}\mathrm{Cov}
\left( X_{\left( 2h-2k\right) },X_{-2l+1}\right) \left( \Psi
_{2l-1}^{0}\right) ^{\ast } \\
&&+\sum_{k\in \mathbb{Z}}\sum_{l\in \mathbb{Z}}\Psi _{2k-1}^{0}\mathrm{Cov}
\left( X_{\left( 2h-2k+1\right) },X_{-2l}\right) \left( \Psi
_{2l}^{0}\right) ^{\ast } \\
&&+\sum_{k\in \mathbb{Z}}\sum_{l\in \mathbb{Z}}\Psi _{2k-1}^{0}\mathrm{Cov}
\left( X_{\left( 2h-2k+1\right) },X_{-2l+1}\right) \left( \Psi
_{2l-1}^{0}\right) ^{\ast }.
\end{eqnarray*}
Consequently,
\begin{eqnarray*}
\mathcal{F}_{\theta ,\left( 0,0\right) }^{\mathbf{Y}} &=&\frac{1}{2\pi }
\sum_{h\in \mathbb{Z}}\mathrm{Cov}\left( \mathbf{Y}_{2h},\mathbf{Y}
_{0}\right) e^{-ih\theta } \\
&=&\frac{1}{2\pi }\sum_{h\in \mathbb{Z}}\sum_{k\in \mathbb{Z}}\sum_{l\in
\mathbb{Z}}\Psi _{2k}^{0}\mathrm{Cov}\left( X_{\left( 2h-2k\right)
},X_{-2l}\right) \left( \Psi _{2l}^{0}\right) ^{\ast }e^{-ih\theta } \\
&&+\frac{1}{2\pi }\sum_{h\in \mathbb{Z}}\sum_{k\in \mathbb{Z}}\sum_{l\in
\mathbb{Z}}\Psi _{2k}^{0}\mathrm{Cov}\left( X_{\left( 2h-2k\right)
},X_{-2l+1}\right) \left( \Psi _{2l-1}^{0}\right) ^{\ast }e^{-ih\theta } \\
&&+\frac{1}{2\pi }\sum_{h\in \mathbb{Z}}\sum_{k\in \mathbb{Z}}\sum_{l\in
\mathbb{Z}}\Psi _{2k-1}^{0}\mathrm{Cov}\left( X_{\left( 2h-2k+1\right)
},X_{-2l}\right) \left( \Psi _{2l}^{0}\right) ^{\ast }e^{-ih\theta } \\
&&+\frac{1}{2\pi }\sum_{h\in \mathbb{Z}}\sum_{k\in \mathbb{Z}}\sum_{l\in
\mathbb{Z}}\Psi _{2k-1}^{0}\mathrm{Cov}\left( X_{\left( 2h-2k+1\right)
},X_{-2l+1}\right) \left( \Psi _{2l-1}^{0}\right) ^{\ast }e^{-ih\theta },
\end{eqnarray*}
which leads
\begin{eqnarray*}
\mathcal{F}_{\theta ,\left( 0,0\right) }^{\mathbf{Y}} &=&\frac{1}{2\pi }
\sum_{k\in \mathbb{Z}}\sum_{l\in \mathbb{Z}}\sum_{h\in \mathbb{Z}}\Psi
_{2k}^{0}\mathrm{Cov}\left( X_{\left( 2h-2k+2l\right) },X_{0}\right) \left(
\Psi _{2l}^{0}\right) ^{\ast }e^{-i\left( h-k+l\right) \theta }e^{il\theta
}e^{-ik\theta } \\
&&+\frac{1}{2\pi }\sum_{k\in \mathbb{Z}}\sum_{l\in \mathbb{Z}}\sum_{h\in
\mathbb{Z}}\Psi _{2k}^{0}\mathrm{Cov}\left( X_{\left( 2h-2k+2l\right)
},X_{1}\right) \left( \Psi _{2l-1}^{0}\right) ^{\ast }e^{-i\left(
h-k+l\right) \theta }e^{il\theta }e^{-ik\theta } \\
&&+\frac{1}{2\pi }\sum_{k\in \mathbb{Z}}\sum_{l\in \mathbb{Z}}\sum_{h\in
\mathbb{Z}}\Psi _{2k-1}^{0}\mathrm{Cov}\left( X_{\left( 2h-2k+2l+1\right)
},X_{0}\right) \left( \Psi _{2l}^{0}\right) ^{\ast }e^{-i\left( h-k+l\right)
\theta }e^{il\theta }e^{-ik\theta } \\
&&+\frac{1}{2\pi }\sum_{k\in \mathbb{Z}}\sum_{l\in \mathbb{Z}}\sum_{h\in
\mathbb{Z}}\Psi _{2k-1}^{0}\mathrm{Cov}\left( X_{\left( 2h-2k+2l+1\right)
},X_{1}\right) \left( \Psi _{2l-1}^{0}\right) ^{\ast }e^{-i\left(
h-k+l\right) \theta }e^{il\theta }e^{-ik\theta },
\end{eqnarray*}
\begin{eqnarray*}
&=&\sum_{k\in \mathbb{Z}}\sum_{l\in \mathbb{Z}}\Psi _{2k}^{0}\mathcal{F}
_{\theta ,\left( 0,0\right) }^{X}\left( \Psi _{2l}^{0}\right) ^{\ast
}e^{il\theta }e^{-ik\theta } \\
&&+\sum_{k\in \mathbb{Z}}\sum_{l\in \mathbb{Z}}\Psi _{2k}^{0}\mathcal{F}
_{\theta ,\left( 0,1\right) }^{X}\left( \Psi _{2l-1}^{0}\right) ^{\ast
}e^{il\theta }e^{-ik\theta } \\
&&+\sum_{k\in \mathbb{Z}}\sum_{l\in \mathbb{Z}}\Psi _{2k-1}^{0}\mathcal{F}
_{\theta ,\left( 1,0\right) }^{X}\left( \Psi _{2l}^{0}\right) ^{\ast
}e^{il\theta }e^{-ik\theta } \\
&&+\sum_{k\in \mathbb{Z}}\sum_{l\in \mathbb{Z}}\Psi _{2k-1}^{0}\mathcal{F}
_{\theta ,\left( 1,1\right) }^{X}\left( \Psi _{2l-1}^{0}\right) ^{\ast
}e^{il\theta }e^{-ik\theta }
\end{eqnarray*}
\begin{eqnarray*}
&=&:\Psi _{\theta ,0}^{0}\mathcal{F}_{\theta ,\left( 0,0\right) }^{X}\left(
\Psi _{\theta ,0}^{0}\right) ^{\ast } \\
&&+\Psi _{\theta ,0}^{0}\mathcal{F}_{\theta ,\left( 0,1\right) }^{X}\left(
\Psi _{\theta ,-1}^{0}\right) ^{\ast } \\
&&+\Psi _{\theta ,-1}^{0}\mathcal{F}_{\theta ,\left( 1,0\right) }^{X}\left(
\Psi _{\theta ,0}^{0}\right) ^{\ast } \\
&&+\Psi _{\theta ,-1}^{0}\mathcal{F}_{\theta ,\left( 1,1\right) }^{X}\left(
\Psi _{\theta ,-1}^{0}\right) ^{\ast }.
\end{eqnarray*}
The operator
$\mathcal{F}_{\theta ,\left( 0,0\right) }^{\mathbf{Y}}$  from $
\mathbb{C}^{p}$ to $\mathbb{C}^{p}$ has the following matrix form
\begin{eqnarray*}
&&\left( \left\langle \left( \mathcal{F}_{\theta ,\left( 0,0\right)
}^{X}\right) ^{\ast }\left( \Psi _{\theta ,0,r}^{0}\right) ,\Psi _{\theta
,0,q}^{0}\right\rangle _{\mathcal{H}}\right) _{p\times p} \\
&&+\left( \left\langle \left( \mathcal{F}_{\theta ,\left( 0,1\right)
}^{X}\right) ^{\ast }\left( \Psi _{\theta ,0,r}^{0}\right) ,\Psi _{\theta
,-1,q}^{0}\right\rangle _{\mathcal{H}}\right) _{p\times p} \\
&&+\left( \left\langle \left( \mathcal{F}_{\theta ,\left( 1,0\right)
}^{X}\right) ^{\ast }\left( \Psi _{\theta ,-1,r}^{0}\right) ,\Psi _{\theta
,0,q}^{0}\right\rangle _{\mathcal{H}}\right) _{p\times p} \\
&&+\left( \left\langle \left( \mathcal{F}_{\theta ,\left( 1,1\right)
}^{X}\right) ^{\ast }\left( \Psi _{\theta ,-1,r}^{0}\right) ,\Psi _{\theta
,-1,q}^{0}\right\rangle _{\mathcal{H}}\right) _{p\times p},
\end{eqnarray*}
Finally,
\begin{eqnarray*}
\mathcal{F}_{\theta ,\left( 0,0\right) }^{\mathbf{Y}} &=&\left\langle \left(
\begin{array}{cc}
\left( \mathcal{F}_{\theta ,\left( 0,0\right) }^{X}\right) ^{\ast } & \left(
\mathcal{F}_{\theta ,\left( 1,0\right) }^{X}\right) ^{\ast } \\
\left( \mathcal{F}_{\theta ,\left( 0,1\right) }^{X}\right) ^{\ast } & \left(
\mathcal{F}_{\theta ,\left( 1,1\right) }^{X}\right) ^{\ast }
\end{array}
\right) \left(
\begin{array}{c}
\Psi _{\theta ,0,r}^{0} \\
\Psi _{\theta ,-1,r}^{0}
\end{array}
\right) ,\left(
\begin{array}{c}
\Psi _{\theta ,0,q}^{0} \\
\Psi _{\theta ,-1,q}^{0}
\end{array}
\right) \right\rangle _{\mathcal{H}^{2}} \\
&=&\left\langle \left(
\begin{array}{cc}
\mathcal{F}_{\theta ,\left( 0,0\right) }^{X} & \mathcal{F}_{\theta ,\left(
0,1\right) }^{X} \\
\mathcal{F}_{\theta ,\left( 1,0\right) }^{X} & \mathcal{F}_{\theta ,\left(
1,1\right) }^{X}
\end{array}
\right) \left(
\begin{array}{c}
\Psi _{\theta ,0,r}^{0} \\
\Psi _{\theta ,-1,r}^{0}
\end{array}
\right) ,\left(
\begin{array}{c}
\Psi _{\theta ,0,q}^{0} \\
\Psi _{\theta ,-1,q}^{0}
\end{array}
\right) \right\rangle _{\mathcal{H}^{2}}.
\end{eqnarray*}
Using similar arguments leads to the following representations for $\mathcal{
F}_{\theta ,\left( 1,0\right) }^{\mathbf{Y}}$, $\mathcal{F}_{\theta ,\left(
0,1\right) }^{\mathbf{Y}}$, and $\mathcal{F}_{\theta ,\left( 1,1\right) }^{
\mathbf{Y}}$.
\begin{equation*}
\mathcal{F}_{\theta ,\left( 1,0\right) }^{\mathbf{Y}}=\left\langle \left(
\begin{array}{cc}
\mathcal{F}_{\theta ,\left( 0,0\right) }^{X} & \mathcal{F}_{\theta ,\left(
0,1\right) }^{X} \\
\mathcal{F}_{\theta ,\left( 1,0\right) }^{X} & \mathcal{F}_{\theta ,\left(
1,1\right) }^{X}
\end{array}
\right) \left(
\begin{array}{c}
\Psi _{\theta ,1,r}^{1} \\
\Psi _{\theta ,0,r}^{1}
\end{array}
\right) ,\left(
\begin{array}{c}
\Psi _{\theta ,0,q}^{0} \\
\Psi _{\theta ,-1,q}^{0}
\end{array}
\right) \right\rangle _{\mathcal{H}^{2}}
\end{equation*}
\begin{equation*}
\mathcal{F}_{\theta ,\left( 0,1\right) }^{\mathbf{Y}}=\left\langle \left(
\begin{array}{cc}
\mathcal{F}_{\theta ,\left( 0,0\right) }^{X} & \mathcal{F}_{\theta ,\left(
0,1\right) }^{X} \\
\mathcal{F}_{\theta ,\left( 1,0\right) }^{X} & \mathcal{F}_{\theta ,\left(
1,1\right) }^{X}
\end{array}
\right) \left(
\begin{array}{c}
\Psi _{\theta ,0,r}^{0} \\
\Psi _{\theta ,-1,r}^{0}
\end{array}
\right) ,\left(
\begin{array}{c}
\Psi _{\theta ,1,q}^{1} \\
\Psi _{\theta ,0,q}^{1}
\end{array}
\right) \right\rangle _{\mathcal{H}^{2}}
\end{equation*}
\begin{equation*}
\mathcal{F}_{\theta ,\left( 1,1\right) }^{\mathbf{Y}}=\left\langle \left(
\begin{array}{cc}
\mathcal{F}_{\theta ,\left( 0,0\right) }^{X} & \mathcal{F}_{\theta ,\left(
0,1\right) }^{X} \\
\mathcal{F}_{\theta ,\left( 1,0\right) }^{X} & \mathcal{F}_{\theta ,\left(
1,1\right) }^{X}
\end{array}
\right) \left(
\begin{array}{c}
\Psi _{\theta ,1,r}^{1} \\
\Psi _{\theta ,0,r}^{1}
\end{array}
\right) ,\left(
\begin{array}{c}
\Psi _{\theta ,1,q}^{1} \\
\Psi _{\theta ,0,q}^{1}
\end{array}
\right) \right\rangle _{\mathcal{H}^{2}}.
\end{equation*}
Note that the 2-periodic behavior of the covariance operators of the filtered
process $\mathbf{Y}$ is an implicit result of the above argument, which
completes the proof of Theorem \ref{T filter process} for the special case $
T=2$. The general case is similar.

\rightline{\QED}

\medskip

\noindent\textsc{Proof of Proposition \ref{p:properties}}: To
establish part (a), consider the eigenvalue decomposition
\eqref{eigenvf decomposition}. We then have
\begin{eqnarray*}
\lambda _{\theta ,m}\varphi _{\theta ,m} &=&\left(
\begin{array}{cc}
\mathcal{F}_{\theta ,\left( 0,0\right) }^{X} & \mathcal{F}_{\theta ,\left(
0,1\right) }^{X} \\
\mathcal{F}_{\theta ,\left( 1,0\right) }^{X} & \mathcal{F}_{\theta ,\left(
1,1\right) }^{X}
\end{array}
\right) \left( \varphi _{\theta ,m}\right) \\
&=&\left(
\begin{array}{cc}
\mathcal{F}_{\theta ,\left( 0,0\right) }^{X} & \mathcal{F}_{\theta ,\left(
0,1\right) }^{X} \\
\mathcal{F}_{\theta ,\left( 1,0\right) }^{X} & \mathcal{F}_{\theta ,\left(
1,1\right) }^{X}
\end{array}
\right) \left(
\begin{array}{c}
\varphi _{\theta ,m,1} \\
\varphi _{\theta ,m,2}
\end{array}
\right) \\
&=&\left(
\begin{array}{c}
\mathcal{F}_{\theta ,\left( 0,0\right) }^{X}\left( \varphi _{\theta
,m,1}\right) +\mathcal{F}_{\theta ,\left( 0,1\right) }^{X}\left( \varphi
_{\theta ,m,2}\right) \\
\mathcal{F}_{\theta ,\left( 1,0\right) }^{X}\left( \varphi _{\theta
,m,1}\right) +\mathcal{F}_{\theta ,\left( 1,1\right) }^{X}\left( \varphi
_{\theta ,m,2}\right)
\end{array}
\right) \\
&=&\frac{1}{2\pi }\sum_{h\in \mathbb{Z}}\left(
\begin{array}{c}
E\left[ \left\langle \varphi _{\theta ,m,1},X_{0}\right\rangle _{\mathcal{H}
}+\left\langle \varphi _{\theta ,m,2},X_{1}\right\rangle _{\mathcal{H}}
\right] X_{2h} \\
E\left[ \left\langle \varphi _{\theta ,m,1},X_{0}\right\rangle _{\mathcal{H}
}+\left\langle \varphi _{\theta ,m,2},X_{1}\right\rangle _{\mathcal{H}}
\right] X_{2h+1}
\end{array}
\right) e^{-ih\theta }.
\end{eqnarray*}
Consequently,
\begin{eqnarray*}
\lambda _{\theta ,m}\overline{\varphi }_{\theta ,m} &=&\frac{1}{2\pi }
\sum_{h\in \mathbb{Z}}\left(
\begin{array}{c}
E\left[ \overline{\left\langle \varphi _{\theta ,m,1},X_{0}\right\rangle }_{
\mathcal{H}}+\overline{\left\langle \varphi _{\theta
,m,2},X_{1}\right\rangle }_{\mathcal{H}}\right] \overline{X}_{2h} \\
E\left[ \overline{\left\langle \varphi _{\theta ,m,1},X_{0}\right\rangle }_{
\mathcal{H}}+\overline{\left\langle \varphi _{\theta
,m,2},X_{1}\right\rangle }_{\mathcal{H}}\right] \overline{X}_{2h+1}
\end{array}
\right) e^{+ih\theta } \\
&&\frac{1}{2\pi }\sum_{h\in \mathbb{Z}}\left(
\begin{array}{c}
E\left[ \left\langle \overline{\varphi }_{\theta ,m,1},X_{0}\right\rangle _{
\mathcal{H}}+\left\langle \overline{\varphi }_{\theta
,m,2},X_{1}\right\rangle _{\mathcal{H}}\right] X_{2h} \\
E\left[ \left\langle \overline{\varphi }_{\theta ,m,1},X_{0}\right\rangle _{
\mathcal{H}}+\left\langle \overline{\varphi }_{\theta
,m,2},X_{1}\right\rangle _{\mathcal{H}}\right] X_{2h+1}
\end{array}
\right) e^{+ih\theta } \\
&=&\left(
\begin{array}{cc}
\mathcal{F}_{-\theta ,\left( 0,0\right) }^{X} & \mathcal{F}_{-\theta ,\left(
0,1\right) }^{X} \\
\mathcal{F}_{-\theta ,\left( 1,0\right) }^{X} & \mathcal{F}_{-\theta ,\left(
1,1\right) }^{X}
\end{array}
\right) \left( \overline{\varphi }_{\theta ,m}\right) .
\end{eqnarray*}
Hence $\lambda _{\theta ,m}$\ and $\overline{\varphi }_{\theta ,m}$ are
an  eigenvalue/eigenfunction pair of $\left(
\begin{array}{cc}
\mathcal{F}_{-\theta ,\left( 0,0\right) }^{X} & \mathcal{F}_{-\theta ,\left(
0,1\right) }^{X} \\
\mathcal{F}_{-\theta ,\left( 1,0\right) }^{X} & \mathcal{F}_{-\theta ,\left(
1,1\right) }^{X}
\end{array}
\right) $. Now, use
\eqref{DF filter}
to obtain
$
\Phi _{l,m}^{t}=\overline{\Phi }_{l,m}^{t},
$
which implies that the DFPC scores $Y_{t,m}$ satisfy
\begin{eqnarray*}
Y_{t,m} &=&\sum_{l\in \mathbb{Z}}\left\langle X_{t-l},\Phi
_{l,m}^{t}\right\rangle _{\mathcal{H}}=\sum_{l\in \mathbb{Z}}\left\langle
\overline{X}_{t-l},\overline{\Phi }_{l,m}^{t}\right\rangle _{\mathcal{H}} \\
&=&\sum_{l\in \mathbb{Z}}\overline{\left\langle X_{t-l},\Phi
_{l,m}^{t}\right\rangle }_{\mathcal{H}}=\overline{Y}_{t,m},
\end{eqnarray*}
and so are real  for each $t$ and $m$.

For part (b),  first we define $Y_{t,m,n}:=\sum_{l=-n}^{n}\left\langle
X_{t-l},\Phi _{l,m}^{t}\right\rangle $. Then we use a similar argument as in
the proof of Theorem \ref{T filter process} to show that $Y_{t,m,n}$\ is
converges in mean-square to $Y_{t,m}=\sum_{l \in \mathbb{Z}}\left\langle
X_{t-l},\Phi _{l,m}^{t}\right\rangle $. Hence
\begin{equation*}
\left\Vert E\left( Y_{t,m,n}\otimes Y_{t,m,n}\right) -E\left( Y_{t,m}\otimes
Y_{t,m}\right) \right\Vert _{\mathcal{S}}\longrightarrow 0,
\text{ \ \textrm{as} }n\longrightarrow \infty,
\end{equation*}
or equevalently,
\begin{equation*}
\left\vert E\left\Vert Y_{t,m,n}\right\Vert _{\mathbb{C}}^{2}-E\left\Vert
Y_{t,m}\right\Vert _{\mathbb{C}}^{2}\right\vert \longrightarrow 0,
\text{ \ \textrm{as} }n\longrightarrow \infty.
\end{equation*}
Consequently, for $t\overset{2}{\equiv }0 $,  we have
\begin{eqnarray*}
E\left\Vert Y_{t,m}\right\Vert _{\mathbb{C}}^{2} &=&\lim_{n\rightarrow
\infty }EY_{t,m,n}\overline{Y}_{t,m,n}=\lim_{n\rightarrow \infty
}\sum_{\left\vert k\right\vert \leq n}\sum_{\left\vert l\right\vert \leq
n}E\left\langle X_{t-l},\Phi _{l,m}^{0}\right\rangle \left\langle \Phi
_{k,m}^{0},X_{t-k}\right\rangle \\
&=&\sum_{k\in \mathbb{Z}}\sum_{l\in \mathbb{Z}}E\left\langle X_{t-2l},\Phi
_{2l,m}^{0}\right\rangle \left\langle \Phi _{2k,m}^{0},X_{t-2k}\right\rangle
\\
&&+\sum_{k\in \mathbb{Z}}\sum_{l\in \mathbb{Z}}E\left\langle X_{t-2l},\Phi
_{2l,m}^{0}\right\rangle \left\langle \Phi
_{2k-1,m}^{0},X_{t-2k+1}\right\rangle \\
&&+\sum_{k\in \mathbb{Z}}\sum_{l\in \mathbb{Z}}E\left\langle X_{t-2l+1},\Phi
_{2l-1,m}^{0}\right\rangle \left\langle \Phi
_{2k,m}^{0},X_{t-2k}\right\rangle \\
&&+\sum_{k\in \mathbb{Z}}\sum_{l\in \mathbb{Z}}E\left\langle X_{t-2l+1},\Phi
_{2l-1,m}^{0}\right\rangle \left\langle \Phi
_{2k-1,m}^{0},X_{t-2k+1}\right\rangle \\
&=&\sum_{k\in \mathbb{Z}}\sum_{l\in \mathbb{Z}}\left\langle
C_{k-l,(0,0)}^{X}\left( \Phi _{2k,m}^{0}\right) ,\Phi
_{2l,m}^{0}\right\rangle _{\mathcal{H}} \\
&&+\sum_{k\in \mathbb{Z}}\sum_{l\in \mathbb{Z}}\left\langle
C_{k-l,(0,1)}^{X}\left( \Phi _{2k-1,m}^{0}\right) ,\Phi
_{2l,m}^{0}\right\rangle _{\mathcal{H}} \\
&&+\sum_{k\in \mathbb{Z}}\sum_{l\in \mathbb{Z}}\left\langle
C_{k-l,(1,0)}^{X}\left( \Phi _{2k,m}^{0}\right) ,\Phi
_{2l-1,m}^{0}\right\rangle _{\mathcal{H}} \\
&&+\sum_{k\in \mathbb{Z}}\sum_{l\in \mathbb{Z}}\left\langle
C_{k-l,(1,1)}^{X}\left( \Phi _{2k-1,m}^{0}\right) ,\Phi
_{2l-1,m}^{0}\right\rangle _{\mathcal{H}}.
\end{eqnarray*}
That is the desired result for the case $t\overset{2}{\equiv }0$.
The case $t\overset{2}{\equiv }1$ is handled in a similar way.

Part (c) is a direct result of Theorem \ref{T filter process}, so  we can
proceed to the proof of part (d). Considering part (c) and using the results
of Proposition 3 of \citetext{hormann:kidzinski:hallin:2015} lead to the
desired result for $2n$ in place of $n$.
\begin{eqnarray*}
&&\lim_{n\rightarrow \infty }\frac{1}{2n}\mathrm{Var}\left( \mathbf{Y}
_{1}+\cdots +\mathbf{Y}_{2n}\right) \\
&=&\lim_{n\rightarrow \infty }\frac{1}{2n}\left[ \mathrm{Var}\left( \mathbf{Y
}_{1}+\mathbf{Y}_{3}+\cdots +\mathbf{Y}_{2n-1}\right) +\mathrm{Var}\left(
\mathbf{Y}_{2}+\mathbf{Y}_{4}+\cdots +\mathbf{Y}_{2n}\right) \right] \\
&=&\dfrac{2\pi}{2} \left[ \mathrm{diag}\left( \lambda _{0,1} ,\ldots
,\lambda _{0,p} \right) +\mathrm{diag}\left( \lambda _{0,p+1} ,\ldots
,\lambda _{0,2p} \right) \right] ,
\end{eqnarray*}
and similarly for $2n+1$ in place of $2n$. This  completes the proof.

\rightline{\QED}

\medskip

\noindent \textsc{Proofs of Theorems \ref{T Inversion Formula} and \ref{T
Optimality}}: Consider the $\mathcal{H}^{2}$-valued mean zero
stationary
process $
\underline{X}=\left\{ \underline{X}_{t}=\left(
\begin{array}{cc}
X_{2t} & X_{2t+1}
\end{array}
\right) ^{\prime },t\in \mathbb{Z}\right\} $ and the filter $\left\{ \mathbf{
\Psi }_{l},l\in \mathbb{Z}\right\} $\ with the following matrix form
\begin{equation*}
\mathbf{\Psi }_{l}=\left(
\begin{array}{cc}
\Psi _{2l}^{0} & \Psi _{2l-1}^{0} \\
\Psi _{2l+1}^{1} & \Psi _{2l}^{1}
\end{array}
\right) :\mathcal{H}^{2}\mathcal{\longrightarrow }\left( \left( \mathbb{C}
^{p}\right) ^{2}\right) \mathbb{C}^{2p},
\end{equation*}
where
\begin{eqnarray*}
\Psi _{l}^{t} &:&\mathcal{H\longrightarrow }\mathbb{C}^{p} \\
\Psi _{l}^{t} &:&h\longmapsto \left( \left\langle h,\Psi
_{l,1}^{t}\right\rangle ,\ldots ,\left\langle h,\Psi _{l,p}^{t}\right\rangle
\right) ^{\prime },\text{\ \ \ \ }t=0,1,\text{\ }l\in \mathbb{Z}.
\end{eqnarray*}
Similarly, define the sequence of operators $\left\{ \mathbf{\Upsilon }
_{l},l\in \mathbb{Z}\right\} $\ with
\begin{equation*}
\mathbf{\Upsilon }_{-l}=\left(
\begin{array}{cc}
\Upsilon _{2l}^{0} & \Upsilon _{2l+1}^{0} \\
\Upsilon _{2l-1}^{1} & \Upsilon _{2l}^{1}
\end{array}
\right) :\left( \left( \mathbb{C}^{p}\right) ^{2}\right) \mathbb{C}^{2p}
\mathcal{\longrightarrow H}^{2},
\end{equation*}
where
\begin{eqnarray*}
\Upsilon _{l}^{t} &:&\mathbb{C}^{p}\mathcal{\longrightarrow H} \\
\Upsilon _{l}^{t} &:&y\longmapsto \sum_{m=1}^{p}y_{m}\Upsilon _{l,m}^{t},
\text{\ \ \ \ }t=0,1,\text{\ }l\in \mathbb{Z}.
\end{eqnarray*}
Therefore,
\begin{equation*}
\mathbf{\Upsilon }\left( B\right) \mathbf{\Psi }\left( B\right) \underline{X}
_{t}=\sum_{m=1}^{p}\left(
\begin{array}{c}
\widetilde{\underline{X}}_{2t,m} \\
\widetilde{\underline{X}}_{2t+1,m}
\end{array}
\right) .
\end{equation*}
On the other hand,  there exist elements $\psi _{q}^{l}=\left(
\begin{array}{cc}
\psi _{q,1}^{l} & \psi _{q,2}^{l}
\end{array}
\right) ^{\prime }$ and $\upsilon _{q}^{l}=\left(
\begin{array}{cc}
\upsilon _{q,1}^{l} & \upsilon _{q,2}^{l}
\end{array}
\right) ^{\prime }$, $q=1,\ldots ,2p$,  in $\mathcal{H}^{2}$, such that
\begin{eqnarray*}
\mathbf{\Psi }_{l}\left( h\right)  &=&\mathbf{\Psi }_{l}\left( \left(
h_{1},h_{2}\right) ^{\prime }\right) =\left( \left\langle h,\psi
_{1}^{l}\right\rangle _{\mathcal{H}^{2}},\ldots ,\left\langle h,\psi
_{2p}^{l}\right\rangle _{\mathcal{H}^{2}}\right) ^{\prime } \\
&=&\left( \left\langle h_{1},\psi _{1,1}^{l}\right\rangle +\left\langle
h_{2},\psi _{1,2}^{l}\right\rangle ,\ldots ,\left\langle h_{1},\psi
_{2p,1}^{l}\right\rangle +\left\langle h_{2},\psi _{2p,2}^{l}\right\rangle
\right) ^{\prime },\text{\ \ \ \ }\forall h_{1},h_{2}\in \mathcal{H},
\end{eqnarray*}
and
\begin{eqnarray*}
\mathbf{\Upsilon }_{-l}\left( y\right)  &=&\mathbf{\Upsilon }_{-l}\left(
\left( y_{1},y_{2}\right) ^{\prime }\right)  \\
&=&\sum_{m=1}^{p}y_{1,m}\upsilon _{m}^{l}+\sum_{m=1}^{p}y_{2,m}\upsilon
_{m+p}^{l},\text{\ \ \ \ }\forall y_{1},y_{2}\in \mathbb{C}^{p}.
\end{eqnarray*}
Simple calculations lead to the following relations, valid
for $ m=1,\ldots ,p$, which play a crucial
role in the remainder of  the proof:
\[
\psi _{m}^{l} =\left(
\begin{array}{c}
\psi _{m,1}^{l} \\
\psi _{m,2}^{l}
\end{array}
\right) =\left(
\begin{array}{c}
\Psi _{2l,m}^{0} \\
\Psi _{2l-1,m}^{0}
\end{array}
\right),\ \ \ \
\psi _{m+p}^{l} =\left(
\begin{array}{c}
\psi _{m+p,1}^{l} \\
\psi _{m+p,2}^{l}
\end{array}
\right) =\left(
\begin{array}{c}
\Psi _{2l+1,m}^{1} \\
\Psi _{2l,m}^{1}
\end{array}
\right),
\]
\[
\upsilon _{m}^{l} =\left(
\begin{array}{c}
\upsilon _{m,1}^{l} \\
\upsilon _{m,2}^{l}
\end{array}
\right) =\left(
\begin{array}{c}
\Upsilon _{2l,m}^{0} \\
\Upsilon _{2l-1,m}^{1}
\end{array}
\right), \ \ \ \
\upsilon _{m+p}^{l} =\left(
\begin{array}{c}
\upsilon _{m+p,1}^{l} \\
\upsilon _{m+p,2}^{l}
\end{array}
\right) =\left(
\begin{array}{c}
\Upsilon _{2l+1,m}^{0} \\
\Upsilon _{2l,m}^{1}
\end{array}
\right).
\]
According to \citetext{hormann:kidzinski:hallin:2015}, we can minimize
\begin{equation*}
E\left\Vert \underline{X}_{t}-\mathbf{\Upsilon }\left( B\right) \mathbf{\Psi
}\left( B\right) \left( \underline{X}_{t}\right) \right\Vert _{\mathcal{H}
^{2}}^{2}
\end{equation*}
by choosing $\upsilon _{\theta ,m}=\sum_{l\in \mathbb{Z}}\upsilon
_{m}^{l}e^{il\theta }=\psi _{\theta ,m}=\sum_{l\in \mathbb{Z}}\psi
_{m}^{l}e^{il\theta }$\ as the $m$-th eigenfunctions of the spectral density
operator $\mathcal{F}_{\theta }^{\underline{X}}$ of the process $\underline{X
}$. Note that $\mathcal{F}_{\theta }^{\underline{X}}$ is nothing other than
\begin{equation*}
\mathcal{F}_{\theta }^{\underline{X}}\left( h\right) =\mathcal{F}_{\theta }^{
\underline{X}}\left( \left(
\begin{array}{c}
h_{1} \\
h_{2}
\end{array}
\right) \right) =\left(
\begin{array}{cc}
\mathcal{F}_{\theta ,\left( 0,0\right) }^{X} & \mathcal{F}_{\theta ,\left(
0,1\right) }^{X} \\
\mathcal{F}_{\theta ,\left( 1,0\right) }^{X} & \mathcal{F}_{\theta ,\left(
1,1\right) }^{X}
\end{array}
\right) \left(
\begin{array}{c}
h_{1} \\
h_{2}
\end{array}
\right) ,\text{\ \ \ \ }\forall h_{1},h_{2}\in \mathcal{H}.
\end{equation*}
This completes the proof.

\rightline{\QED}

\medskip

\noindent \textsc{Proof of Theorem~\ref{T Consistency}}:
We will use the continuity of each function $\lambda_{\cdot,m}$.
This is a direct consequence of
Proposition 7 of \citetext{hormann:kidzinski:hallin:2015}.
As in the proof of
Theorem 3 of \citetext{hormann:kidzinski:hallin:2015},
we show that $E\left\vert
Y_{t,m}-\widehat{Y}_{t.m}\right\vert \longrightarrow 0$.
For a fixed $L$ and $t
\overset{2}{\equiv }0$\ we have,
\begin{eqnarray}
E\left\vert Y_{t,m}-\widehat{Y}_{t.m}\right\vert &=&E\left\vert
\sum_{l=-2L-1}^{2L}\left\langle X_{t-l},\Phi _{l,m}^{0}-\widehat{\Phi }
_{l,m}^{0}\right\rangle _{\mathcal{H}}+\sum_{l\notin \left[ -2L-1,2L\right]
}\left\langle X_{t-l},\Phi _{l,m}^{0}\right\rangle _{\mathcal{H}}\right\vert
\notag \\
&\leq &E\left\vert \sum_{l=-2L-1}^{2L}\left\langle X_{t-l},\Phi _{l,m}^{0}-
\widehat{\Phi }_{l,m}^{0}\right\rangle _{\mathcal{H}}\right\vert
+E\left\vert \sum_{l\notin \left[ -2L-1,2L\right] }\left\langle X_{t-l},\Phi
_{l,m}^{0}\right\rangle _{\mathcal{H}}\right\vert
\label{two converging terms to zero} \\
&=&E\left\vert \sum_{l=-L}^{L}\left\langle X_{t-2l},\Phi _{2l,m}^{0}-
\widehat{\Phi }_{2l,m}^{0}\right\rangle _{\mathcal{H}}+\sum_{l=-L}^{L}\left
\langle X_{t-2l+1},\Phi _{2l-1,m}^{0}-\widehat{\Phi }_{2l-1,m}^{0}\right
\rangle _{\mathcal{H}}\right\vert  \notag \\
&&+E\left\vert \sum_{\left\vert l\right\vert >L}\left\langle X_{t-2l},\Phi
_{2l,m}^{0}\right\rangle _{\mathcal{H}}+\sum_{\left\vert l\right\vert
>L}\left\langle X_{t-2l+1},\Phi _{2l-1,m}^{0}\right\rangle _{\mathcal{H}
}\right\vert .  \notag
\end{eqnarray}
Thus,  it is enough to show that each of the terms appearing in
\eqref{two converging terms to zero}
converges to zero in probability. The first term can be bounded as follows:
\begin{eqnarray}
&&E\left\vert \sum_{l=-2L-1}^{2L} \left\langle X_{t-l},\Phi _{l,m}^{0}-
\widehat{\Phi }_{l,m}^{0}\right\rangle _{\mathcal{H}}\right\vert  \notag \\
&\leq &E\sum_{l=-2L-1}^{2L}\left\Vert X_{t-l}\right\Vert \left\Vert \Phi
_{l,m}^{0}-\widehat{\Phi }_{l,m}^{0}\right\Vert  \notag \\
&=&E\sum_{l=-L}^{L}\left\Vert X_{t-2l}\right\Vert \left\Vert \Phi
_{2l,m}^{0}-\widehat{\Phi }_{2l,m}^{0}\right\Vert  \notag \\
&&+E\sum_{l=-L}^{L}\left\Vert X_{t-2l+1}\right\Vert \left\Vert \Phi
_{2l-1,m}^{0}-\widehat{\Phi }_{2l-1,m}^{0}\right\Vert  \notag \\
&\leq &E\sum_{l=-L}^{L}\left\Vert X_{t-2l}\right\Vert \left( \left\Vert \Phi
_{2l,m}^{0}-\widehat{\Phi }_{2l,m}^{0}\right\Vert ^{2}+\left\Vert \Phi
_{2l-1,m}^{0}-\widehat{\Phi }_{2l-1,m}^{0}\right\Vert ^{2}\right) ^{\frac{1}{
2}}  \notag \\
&&+E\sum_{l=-L}^{L}\left\Vert X_{t-2l+1}\right\Vert \left( \left\Vert \Phi
_{2l,m}^{0}-\widehat{\Phi }_{2l,m}^{0}\right\Vert ^{2}+\left\Vert \Phi
_{2l-1,m}^{0}-\widehat{\Phi }_{2l-1,m}^{0}\right\Vert ^{2}\right) ^{\frac{1}{
2}}  \notag \\
&\leq &2E\sum_{l=-L}^{L}\left\Vert \underline{X}_{\left( \frac{t}{2}
-l\right) }\right\Vert _{\mathcal{H}^{2}}\left\Vert \left(
\begin{array}{cc}
\Phi _{2l,m}^{0} & \Phi _{2l-1,m}^{0}
\end{array}
\right) ^{\prime }-\left(
\begin{array}{cc}
\widehat{\Phi }_{2l,m}^{0} & \widehat{\Phi }_{2l-1,m}^{0}
\end{array}
\right) ^{\prime }\right\Vert _{\mathcal{H}^{2}}.  \label{consistency part I}
\end{eqnarray}
Next, we use \eqref{DF filter} to obtain
\begin{eqnarray*}
&&2\pi \left\Vert \left(
\begin{array}{cc}
\Phi _{2l,m}^{0} & \Phi _{2l-1,m}^{0}
\end{array}
\right) ^{\prime }-\left(
\begin{array}{cc}
\widehat{\Phi }_{2l,m}^{0} & \widehat{\Phi }_{2l-1,m}^{0}
\end{array}
\right) ^{\prime }\right\Vert _{\mathcal{H}^{2}} \\
&=&\left\Vert \int_{-\pi }^{\pi }\left( \varphi _{\theta ,m}-\widehat{
\varphi }_{\theta ,m}\right) e^{-il\theta }d\theta \right\Vert _{\mathcal{H}
^{2}} \\
&\leq &\int_{-\pi }^{\pi }\left\Vert \varphi _{\theta ,m}-\widehat{\varphi }
_{\theta ,m}\right\Vert _{\mathcal{H}^{2}}d\theta \\
&=&\int_{-\pi }^{\pi }\left\Vert \varphi _{\theta ,m}-\left( 1-\widehat{c}
_{\theta ,m}+\widehat{c}_{\theta ,m}\right) \widehat{\varphi }_{\theta
,m}\right\Vert _{\mathcal{H}^{2}}d\theta \\
&\leq &\int_{-\pi }^{\pi }\left\Vert \varphi _{\theta ,m}-\widehat{c}
_{\theta ,m}\widehat{\varphi }_{\theta ,m}\right\Vert _{\mathcal{H}
^{2}}d\theta +\int_{-\pi }^{\pi }\left\Vert \left( 1-\widehat{c}_{\theta
,m}\right) \widehat{\varphi }_{\theta ,m}\right\Vert _{\mathcal{H}
^{2}}d\theta \\
&=&\int_{-\pi }^{\pi }\left\Vert \varphi _{\theta ,m}-\widehat{c}_{\theta ,m}
\widehat{\varphi }_{\theta ,m}\right\Vert _{\mathcal{H}^{2}}d\theta
+\int_{-\pi }^{\pi }\left\vert 1-\widehat{c}_{\theta ,m}\right\vert d\theta
\\
&=&:Q_{1}+Q_{2},
\end{eqnarray*}
where $\widehat{c}_{\theta ,m}:=\frac{\left\langle \varphi _{\theta ,m},
\widehat{\varphi }_{\theta ,m}\right\rangle _{\mathcal{H}^{2}}}{\left\vert
\left\langle \varphi _{\theta ,m},\widehat{\varphi }_{\theta
,m}\right\rangle _{\mathcal{H}^{2}}\right\vert }$, $m=1,\ldots ,p$.
According to Lemma 3.2 of \citetext{hormann:kokoszka:2010} we have the
following upper bound for $Q_{1}$:
\begin{eqnarray}
Q_{1} &\leq &\int_{-\pi }^{\pi }\frac{2\sqrt{2}}{\alpha _{\theta ,m}}
\left\Vert \mathcal{F}_{\theta }^{\underline{X}}-\widehat{\mathcal{F}}
_{\theta }^{\underline{X}}\right\Vert _{\mathcal{L}}\wedge 2d\theta  \notag
\\
&\leq &\int_{-\pi }^{\pi }\frac{2\sqrt{2}}{\alpha _{\theta ,m}}\left\Vert
\mathcal{F}_{\theta }^{\underline{X}}-\widehat{\mathcal{F}}_{\theta }^{
\underline{X}}\right\Vert _{\mathcal{S}}\wedge 2d\theta .
\label{upper bound  for Q_1}
\end{eqnarray}
By Condition \ref{distinct eigenvalues} there is a finite subset $\left\{
\theta _{1},\ldots ,\theta _{K}\right\} $\ of $\left( -\pi ,\pi \right] $\
for which $\alpha _{\theta _{1},m}=\cdots =\alpha _{\theta _{K},m}=0$.
Define $A\left( m,\epsilon \right) :=\bigcup_{j=1}^{K}\left[ \theta
_{j}-\epsilon ,\theta _{j}+\epsilon \right] $ and $M_{\epsilon }^{-1}:=\min
\left\{ \alpha _{\theta ,m}:\theta \in \left[ -\pi ,\pi \right] \setminus
A\left( m,\epsilon \right) \right\} $. Therefore the upper bound
\eqref{upper bound  for Q_1} satisfies
\begin{eqnarray*}
\int_{-\pi }^{\pi }\frac{2\sqrt{2}}{\alpha _{\theta ,m}}\left\Vert \mathcal{F
}_{\theta }^{\underline{X}}-\widehat{\mathcal{F}}_{\theta }^{\underline{X}
}\right\Vert _{\mathcal{S}}\wedge 2d\theta &\leq &4K\epsilon +8M_{\epsilon
}^{2}\int_{-\pi }^{\pi }\frac{2\sqrt{2}}{\alpha _{\theta ,m}}\left\Vert
\mathcal{F}_{\theta }^{\underline{X}}-\widehat{\mathcal{F}}_{\theta }^{
\underline{X}}\right\Vert _{\mathcal{S}}d\theta \\
&:&=B_{n,\epsilon }.
\end{eqnarray*}
Now, use the countinuty of $\lambda _{\cdot ,m}$ and Condition \ref{Consistency
of F hat s} and choose $\epsilon >0$, small enough to conclude $
B_{n,\epsilon }$\ tends to zero in probability. Thus,
\begin{equation}
\int_{-\pi }^{\pi }\left\vert \left\langle \varphi _{\theta ,m},\omega
\right\rangle -\widehat{c}\left( \theta \right) \left\langle \widehat{
\varphi }_{\theta ,m},\omega \right\rangle \right\vert d\theta
\longrightarrow 0\text{ in probability.}  \label{<.,w>}
\end{equation}
By applying similar argument as in  the proof of Theorem 3 of
\citetext{hormann:kidzinski:hallin:2015},  we also conclude that $Q_{2}$\
tends to zero in probability. Remark \ref{finiteness of E||X_t||^2} entails
\begin{eqnarray*}
&&E\left\vert \sum_{l=-2L-1}^{2L}\left\langle X_{t-l},\Phi _{l,m}^{0}-
\widehat{\Phi }_{l,m}^{0}\right\rangle _{\mathcal{H}}\right\vert \\
&\leq &E\sum_{l=-2L-1}^{2L}\left\Vert X_{t-l}\right\Vert _{\mathcal{H}
}\left\Vert \Phi _{l,m}^{0}-\widehat{\Phi }_{l,m}^{0}\right\Vert _{\mathcal{H
}} \\
&\leq &\dfrac{Q_{1}+Q_{2}}{2\pi }\sum_{l=-2L+1}^{2L}E\left\Vert
X_{t-l}\right\Vert _{\mathcal{H}} \\
&\leq &o_{P}\left( 1\right) \frac{\sum_{l=-2L+1}^{2L}E\left\Vert
X_{t-l}\right\Vert _{\mathcal{H}}}{L} \\
&\leq &o_{P}\left( 1\right) \frac{\sum_{l=-2L+1}^{2L}\left( E\left\Vert
X_{t-l}\right\Vert _{\mathcal{H}}^{2}\right) ^{\frac{1}{2}}}{L} \\
&\leq &o_{P}\left( 1\right) \dfrac{(4L+2)\left( \max_{0\leq t\leq T-1}\left(
E\left\Vert X_{t}\right\Vert ^{2}\right) \right) ^{\frac{1}{2}}}{L}
\longrightarrow 0.
\end{eqnarray*}
It remains to show that $E\left\vert \sum_{\left\vert l\right\vert
>L}\left\langle X_{t-2l},\Phi _{2l,m}^{0}\right\rangle _{\mathcal{H}
}+\sum_{\left\vert l\right\vert >L}\left\langle X_{t-2l+1},\Phi
_{2l-1,m}^{0}\right\rangle _{\mathcal{H}}\right\vert ^{2}$ tends to zero.
\begin{eqnarray*}
&&E\left\langle \sum_{\left\vert l\right\vert >L}\left\langle X_{t-2l},\Phi
_{2l,m}^{0}\right\rangle _{\mathcal{H}}+\left\langle X_{t-2l+1},\Phi
_{2l-1,m}^{0}\right\rangle _{\mathcal{H}},\sum_{\left\vert k\right\vert
>L}\left\langle X_{t-2k},\Phi _{2k,m}^{0}\right\rangle _{\mathcal{H}
}+\left\langle X_{t-2k+1},\Phi _{2k-1,m}^{0}\right\rangle _{\mathcal{H}
}\right\rangle _{\mathbb{C}} \\
&=&\sum_{\left\vert l\right\vert >L}\sum_{\left\vert k\right\vert
>L}\left\langle E\left( X_{t-2l}\otimes X_{t-2k}\right) \Phi
_{2k,m}^{0},\Phi _{2l,m}^{0}\right\rangle _{\mathcal{H}} \\
&&+\sum_{\left\vert l\right\vert >L}\sum_{\left\vert k\right\vert
>L}\left\langle E\left( X_{t-2l}\otimes X_{t-2k+1}\right) \Phi
_{2k-1,m}^{0},\Phi _{2l,m}^{0}\right\rangle _{\mathcal{H}} \\
&&+\sum_{\left\vert l\right\vert >L}\sum_{\left\vert k\right\vert
>L}\left\langle E\left( X_{t-2l+1}\otimes X_{t-2k}\right) \Phi
_{2k,m}^{0},\Phi _{2l-1,m}^{0}\right\rangle _{\mathcal{H}} \\
&&+\sum_{\left\vert l\right\vert >L}\sum_{\left\vert k\right\vert
>L}\left\langle E\left( X_{t-2l+1}\otimes X_{t-2k+1}\right) \Phi
_{2k-1,m}^{0},\Phi _{2l-1,m}^{0}\right\rangle _{\mathcal{H}}
\end{eqnarray*}
Cauchy-Schwartz inequality entails
\begin{eqnarray*}
&&E\left\vert \sum_{\left\vert l\right\vert >L}\left\langle X_{t-2l},\Phi
_{2l,m}^{0}\right\rangle _{\mathcal{H}}+\sum_{\left\vert l\right\vert
>L}\left\langle X_{t-2l+1},\Phi _{2l-1,m}^{0}\right\rangle _{\mathcal{H}
}\right\vert ^{2} \\
&\leq &\sum_{\left\vert k\right\vert >L}\sum_{\left\vert l\right\vert
>L}\left\Vert C_{k-l,\left( 0,0\right) }^{X}\right\Vert _{\mathcal{L}
}\left\Vert \Phi _{2k,m}^{0}\right\Vert \left\Vert \Phi
_{2l,m}^{0}\right\Vert \\
&&+\sum_{\left\vert k\right\vert >L}\sum_{\left\vert l\right\vert
>L}\left\Vert C_{k-l,\left( 0,1\right) }^{X}\right\Vert _{\mathcal{L}
}\left\Vert \Phi _{2k-1,m}^{0}\right\Vert \left\Vert \Phi
_{2l,m}^{0}\right\Vert \\
&&+\sum_{\left\vert k\right\vert >L}\sum_{\left\vert l\right\vert
>L}\left\Vert C_{k-l,\left( 1,0\right) }^{X}\right\Vert _{\mathcal{L}
}\left\Vert \Phi _{2k,m}^{0}\right\Vert \left\Vert \Phi
_{2l-1,m}^{0}\right\Vert \\
&&+\sum_{\left\vert k\right\vert >L}\sum_{\left\vert l\right\vert
>L}\left\Vert C_{k-l,\left( 1,1\right) }^{X}\right\Vert _{\mathcal{L}
}\left\Vert \Phi _{2k-1,m}^{0}\right\Vert \left\Vert \Phi
_{2l-1,m}^{0}\right\Vert \\
&=&\sum_{h\in \mathbb{Z}}\sum_{k\in \mathbb{Z}}\left\Vert C_{h,\left(
0,0\right) }^{X}\right\Vert _{\mathcal{L}}\left\Vert \Phi
_{2k,m}^{0}\right\Vert \left\Vert \Phi _{2\left( k-h\right)
,m}^{0}\right\Vert I\left\{ \left\vert k\right\vert >L\right\} I\left\{
\left\vert k-h\right\vert >L\right\} \\
&&+\sum_{h\in \mathbb{Z}}\sum_{k\in \mathbb{Z}}\left\Vert C_{h,\left(
0,1\right) }^{X}\right\Vert _{\mathcal{L}}\left\Vert \Phi
_{2k-1,m}^{0}\right\Vert \left\Vert \Phi _{2\left( k-h\right)
,m}^{0}\right\Vert I\left\{ \left\vert k\right\vert >L\right\} I\left\{
\left\vert k-h\right\vert >L\right\} \\
&&+\sum_{h\in \mathbb{Z}}\sum_{k\in \mathbb{Z}}\left\Vert C_{h,\left(
1,0\right) }^{X}\right\Vert _{\mathcal{L}}\left\Vert \Phi
_{2k,m}^{0}\right\Vert \left\Vert \Phi _{2\left( k-h\right)
-1,m}^{0}\right\Vert I\left\{ \left\vert k\right\vert >L\right\} I\left\{
\left\vert k-h\right\vert >L\right\} \\
&&+\sum_{h\in \mathbb{Z}}\sum_{k\in \mathbb{Z}}\left\Vert C_{h,\left(
1,1\right) }^{X}\right\Vert _{\mathcal{L}}\left\Vert \Phi
_{2k-1,m}^{0}\right\Vert \left\Vert \Phi _{2\left( k-h\right)
-1,m}^{0}\right\Vert I\left\{ \left\vert k\right\vert >L\right\} I\left\{
\left\vert k-h\right\vert >L\right\}
\end{eqnarray*}
\begin{eqnarray*}
&\leq &\sum_{h\in \mathbb{Z}}\left\Vert C_{h,\left( 0,0\right)
}^{X}\right\Vert _{\mathcal{L}}\left( \sum_{k\in \mathbb{Z}}\left\Vert \Phi
_{2k,m}^{0}\right\Vert ^{2}I\left\{ \left\vert k\right\vert >L\right\}
\right) ^{\frac{1}{2}}\left( \sum_{k\in \mathbb{Z}}\left\Vert \Phi
_{2k,m}^{0}\right\Vert ^{2}\right) ^{\frac{1}{2}} \\
&&+\sum_{h\in \mathbb{Z}}\left\Vert C_{h,\left( 0,1\right) }^{X}\right\Vert
_{\mathcal{L}}\left( \sum_{k\in \mathbb{Z}}\left\Vert \Phi
_{2k-1,m}^{0}\right\Vert ^{2}I\left\{ \left\vert k\right\vert >L\right\}
\right) ^{\frac{1}{2}}\left( \sum_{k\in \mathbb{Z}}\left\Vert \Phi
_{2k,m}^{0}\right\Vert ^{2}\right) ^{\frac{1}{2}} \\
&&+\sum_{h\in \mathbb{Z}}\left\Vert C_{h,\left( 1,0\right) }^{X}\right\Vert
_{\mathcal{L}}\left( \sum_{k\in \mathbb{Z}}\left\Vert \Phi
_{2k,m}^{0}\right\Vert ^{2}I\left\{ \left\vert k\right\vert >L\right\}
\right) ^{\frac{1}{2}}\left( \sum_{k\in \mathbb{Z}}\left\Vert \Phi
_{2k-1,m}^{0}\right\Vert ^{2}\right) ^{\frac{1}{2}} \\
&&+\sum_{h\in \mathbb{Z}}\left\Vert C_{h,\left( 1,1\right) }^{X}\right\Vert
_{\mathcal{L}}\left( \sum_{k\in \mathbb{Z}}\left\Vert \Phi
_{2k-1,m}^{0}\right\Vert ^{2}I\left\{ \left\vert k\right\vert >L\right\}
\right) ^{\frac{1}{2}}\left( \sum_{k\in \mathbb{Z}}\left\Vert \Phi
_{2k-1,m}^{0}\right\Vert ^{2}\right) ^{\frac{1}{2}},
\end{eqnarray*}
which obviously tends to zero as $n$\ tends to infinity. Using a similar
argument for case $t\overset{2}{\equiv }1$\ completes the proof.

\rightline{\QED}

\renewcommand{\baselinestretch}{0.9}
\small
\bibliographystyle{oxford3}
\bibliography{neda}

\end{document}